\documentclass[%
preprintnumbers,
twocolumn,
superscriptaddress,
 amsmath,amssymb,
 aps,
prd,
]{revtex4-1}
\usepackage[dvipsnames,usenames]{xcolor}
\usepackage{graphicx}
\usepackage{dcolumn}
\usepackage{bm}
\usepackage{mathrsfs,natbib}
\usepackage{graphicx,epsf,amssymb,amsbsy,amsfonts,amssymb,amsmath}
\usepackage{slashed}
\usepackage{comment}
\usepackage{ulem}
\usepackage{hyperref}
\usepackage[mathlines]{lineno}

\usepackage{epstopdf}

\newcommand{\be}{\begin{equation}}
\newcommand{\ee}{\end{equation}}

\newcommand{\MeV}{{\rm ~MeV }}

\newcommand{\ud}{\mathrm{d}}

\newcommand{\eps}{\epsilon}
\newcommand{\lra}[1]{\langle #1 \rangle}

\begin{document}

\preprint{INT-PUB-18-010}

\title{Dark halos around neutron stars and gravitational waves}
\author{Ann E. Nelson}
\email{aenelson@uw.edu}
\affiliation{
Department of Physics, University of Washington, Seattle, WA 98195
}%
\author{Sanjay Reddy}
\email{sareddy@uw.edu}
\affiliation{
Institute for Nuclear Theory, University of Washington, Seattle, WA 98195
}
\affiliation{
Department of Physics, University of Washington, Seattle, WA 98195
}

\author{Dake Zhou}
\email{zdk@uw.edu}
\affiliation{
Institute for Nuclear Theory, University of Washington, Seattle, WA 98195
}%
\affiliation{
Department of Physics, University of Washington, Seattle, WA 98195
}%
\date{\today}

\begin{abstract}
We find that models of MeV-GeV dark matter in which dark matter interacts strongly can be constrained by the observation of gravitational waves from binary neutron star (BNS) mergers. Trace amounts of dark matter, either produced during the supernova or accreted later, can alter the structure of neutron stars (NS) and influence their tidal deformability.  We focus on models of dark matter interacting by the exchange of light vector gauge bosons that couple to a conserved dark charge. In these models, dark matter accumulated in neutron stars can extend to large radii. Gravitational waves detected from the first observed BNS merger GW170817 places useful constraints on such not-so compact objects. Dark halos, if present, also predict a greater variability of neutron star tidal deformabilities than expected for ordinary neutron stars.
\end{abstract}


\maketitle

\section{Introduction}
The particle nature of dark matter (DM) remains elusive despite developments in detection methods  \cite{Bauer:2013ihz}. Traditionally, indirect detection has focused on evidence for the annihilation products of DM, but there is growing understanding that properties of astrophysical objects might be affected by DM, and that DM might not self-annihilate. The similar abundances of baryons and DM have led to theorizing that DM might, like ordinary matter, carry a conserved charge. It could then be  produced in a similar manner to baryons, by creating an asymmetry between DM and anti DM. This type of DM is known as Asymmetric Dark Matter (ADM). Such an asymmetry would allow for sizable self-interactions without annihilations. While the paradigm of Cold Dark Matter fits a large range of data, some discrepancies at scales smaller than galaxy clusters could be addressed by a more complicated DM sector including large self-interactions \cite{Spergel:2000,Loeb:2011,Davoudiasl:2012uw,Zavala_2013,Bertoni:2014mva,Petraki:2014uza,Buckley:2014hja,Creasey:2016jaq,DiCintio:2017zdz,Tulin:2017ara,Duerr:2018mbd}. Some recent work suggests that a strong velocity dependence in the interaction, such as would be the case if DM exchanges a light boson, provides the best fit to a range of galactic structures \cite{Kaplinghat:2015aga}. If neutron stars trap DM, the resulting DM density is much higher and relative velocity much lower than found in other places in the universe, providing a unique regime to constrain any dark forces. Here we will show how a dark halo surrounding NS resulting from repulsive dark forces can affect gravitational wave production in neutron star mergers.

Interesting ideas about the role of DM inside NSs have been explored and are reviewed in \cite{Kaplan_2009,Boucenna:2013wba,Zurek:2014}. Recent work in \cite{Kouvaris:2010,Leung_2011,Kouvaris_2015,Bertoni:2013bsa} has shown that ADM can accrete and thermalize in the NS core even when the DM-nucleon cross section  $\sigma_{\chi n}$ is significantly smaller than the current terrestrial bound.  The accreted DM mass in a NS is
\begin{equation}
M_{\rm acc} \approx 10^{-16}  \left(\frac{\rho_\chi}{\rm GeV/cm^3}  \right) \left(\frac{\sigma_{\chi n}}{10^{-45} {\rm cm}^2}  \right) \left(\frac{\rm t}{10^8~{\rm yrs}}  \right)~M_\odot\,,
\end{equation}
 where $t$ is the age of the neutron star and $\rho_\chi$ is the ambient density of DM. When the DM mass in the core exceeds a critical value its self-gravity dominates, and in the absence of repulsive interactions bosonic DM becomes unstable and collapses to form a black hole. This black hole is then expected to devour the NS on a short timescale \cite{Goldman:1989nd,Gould:1989gw,Kouvaris:2010,Leung_2011,Kouvaris_2015}. For MeV-GeV asymmetric bosonic DM a Bose Einstein condensate can form earlier and hasten black hole formation. This may happen when the accreted number exceeds $N_c > 10^{36} ~(T/10^5 K)^3$ where $T\simeq 10^5-10^6$ K  is the expected core temperature of an old NS \cite{Jamison:2013yya}. The corresponding critical mass $M_c = N_c m_\chi \simeq 10^{-21} (m_\chi/{\rm GeV})~M_\odot$ is small suggesting that old neutron stars in our galaxy with ages $\simeq 10^8$ yrs should have accreted the DM mass needed to become unstable \cite{Kouvaris:2010}.  Old neutron stars such as J0437-4715 with estimated age $t_{NS}\simeq 10^9$ yrs thereby provide strong constraints on models of weakly interacting asymmetric bosonic DM \cite{McDermott:2011jp}.

Fermionic or bosonic DM with repulsive interactions can evade these constraints and form a stable hydrostatic configuration inside the NS \cite{Bell:2013xk}. Here we consider a simple renormalizable model of DM with self-repulsion. The repulsion arises from the exchange of a vector mediator from a spontaneously broken U(1) gauge interaction. We also couple this vector to Standard Model baryon number. The DM particle which we shall label $\chi$ can be either a charged scalar or a charged Dirac fermion. The model is defined by the following parameters: (i) $m_\chi$ is the mass of the charged fermion or scalar; (ii) $m_\phi$ is the mass of the mediator gauge boson $\phi$; (iii) $g_{\chi}$ is the coupling strength of the gauge boson to the dark charge; and (iv) $g_{B}$, the gauge boson coupling to baryon number, is constrained by experiment and SN 1987a to be $\lesssim 10^{-10}$ \cite{Rrapaj:2015wgs}.

One objective of this study is to identify NS observables that can shed light on interacting DM candidates.  Another is to explore the diversity of gravitational wave (GW) signals from merging NSs which may contain DM.  The possibility of hybrid compact objects containing DM has been discussed earlier in Ref.~\cite{Leung_2011} and implications of a dark core with mass $\simeq10\% $ of the NS mass for post-merger dynamics was explored in Ref.~\cite{Ellis:2017jgp}. Here we calculate for the first time the tidal deformability of hybrid stars to address how trace amounts of DM would form a halo and influence GW emission during inspiral.

The quadrupole polarizability or tidal deformability $\lambda$ of a star is defined by $Q_{ij}=-\lambda E_{ij}$ where $Q_{ij}$ is the induced quadrupole moment  and $E_{ij}$ is the applied tidal field.  It is known that the tidal deformation of ordinary NSs with radii in the range $10-14$ km  can produce discernible changes to the observed waveforms by accelerating inspiral and advancing the phase of the emitted GWs when the orbital radius is larger than but comparable to the radius of NS \cite{Flanagan:2007ix}.

The detection and analysis of GW emission from GW170817 suggests that the dimensionless measure of the tidal deformability  $\Lambda=\lambda/M^5$ where $M$ is the mass of the NS is small. Here, and throughout this letter we set $c=1$ and gravitational constant $G=1$. GW170817 provides an upper bound on tidal deformability of a canonical neutron star with mass $M=1.4~M_\odot$ and requires that  $\Lambda_{1.4M_\odot} < 800$ \cite{TheLIGOScientific:2017qsa} and the 90\% confidence interval $\Lambda_{1.4M_\odot} = 190^{+390}_{-120}$ favors even smaller values \cite{Abbott:2018exr}. Most nuclear physics based equations of state (EOS) predict $\Lambda_{1.4M_\odot}$  in this range and a large parameter space of the DM model defined earlier can be excluded if the NS contain a trace amount of DM with mass $\gtrsim 10^{-5}~M_\odot$.  This is our main result and will be discussed in detail in section \ref{sec:tidal}. In section \ref{sec:acc} we discuss the plausibility of finding trace amounts of DM inside neutron stars, and we conclude in section \ref{sec:con} by providing constraints for the DM interaction strengths for MeV-GeV scale DM.

\section{Dark halos and their tidal deformability}
\label{sec:tidal}
\begin{figure}[!htbp]
\centering
\includegraphics[width=\linewidth]{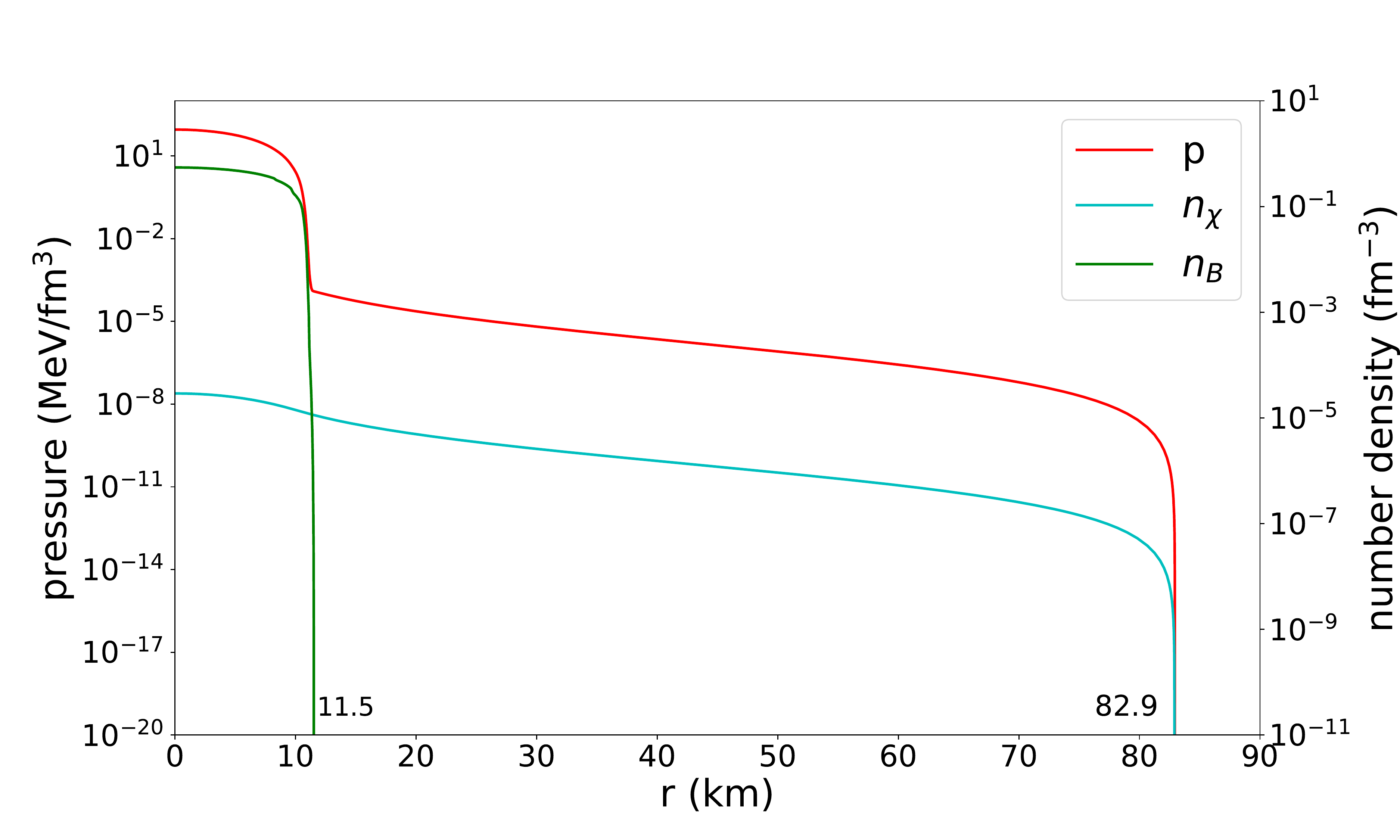}
\caption{Density profile for a hybrid star with total mass $1.4M_\odot$ containing $M_\chi=1.7\times10^{-4}~M_\odot$ of DM. The presence of DM increases the tidal deformability of the NS from $\Lambda^\text{APR}_{1.4M_\odot}=260$ to $\Lambda_{1.4M_\odot}=800$ for $m_\chi=100$ MeV and $g_\chi/m_\phi=5\times10^{-1}$ MeV$^{-1}$. The baryon density is shown in green, the total pressure shown in red, and the dark matter density is depicted by the cyan curve. The radius of dark halo increases with increasing coupling strength $g_\chi$ or decreasing DM particle mass $m_\chi$. The baryon profile is almost independent of DM properties as long as $M_\chi \ll M_{NS}$.}
	\label{fig:profiles}
\end{figure}

As noted earlier fermionic DM and bosonic DM with repulsive interactions can form stable hydrostatic configurations since they have adequate pressure to balance gravity. In this section we shall assume that the DM mass inside the NS  $M_\chi \ll M_{NS}$ where $M_{NS} \simeq 1.4 M_\odot$ is the mass of the neutron star. Since DM will represent a small perturbation of the radial distribution of baryon matter  when $M_\chi$ is small, we first construct a NS composed of only nuclear matter. This is done by choosing a nuclear EOS which specifies the relationship between pressure ($p$) and energy density ($\epsilon$) and solving Tolman-Oppenheimer-Volkoff (TOV) equations for the hydrostatic structure\cite{Tolman:1939jz,Oppenheimer:1939ne}. We adopt a widely used EOS for NS called the Akmal-Pandharipande-Ravenhall (APR) EOS \cite{Akmal:1998cf} to determine the NS structure. Later we will explore a more general parameterizations of the nuclear EOS to account for uncertainties associated with the dense matter EOS.  For a chosen central energy density the solution to TOV equations determine the mass $M$, radius $R$, and radial profiles of the baryon density $n_B(r)$, the energy density $\epsilon(r)$, the pressure $p(r)$, and metric functions $g_{rr}(r)$ and $g_{tt}(r)$ that define the spherically symmetric geometry. The profile of the pressure and the baryon number density inside a NS of mass $M=1.4~M_\odot$ are shown in Fig.~\ref{fig:profiles}.

To include DM we first note that in equilibrium the chemical potential of DM, denoted as $\mu_{\chi}$, should be a constant, and in the presence of neutron star's gravitational field this requires
\begin{equation}
\mu_\chi=\tilde{\mu}_{\chi} (r)~\sqrt{g_{tt}(r)}={\rm constant}\,,
\label{eq:muConstant}
\end{equation}
where $\tilde{\mu}_{\chi} (r)$ is DM chemical potential in local Lorentz frame. If one neglects the back reaction of  DM on the NS structure, a specification of $\tilde{\mu}_{\chi} (r=0)$ uniquely determines the distribution of DM inside the NS.  The number density of DM $n_\chi$ is obtained by noting that $\tilde{\mu}_{\chi} (r)=(\partial \epsilon_\chi/\partial n_\chi)$ where $\epsilon_\chi$ is the DM energy density. Neglecting finite temperature effects, since thermal energies in NS are small,  the energy density of DM is given by (setting $\hbar=1$)
\begin{equation}
 \epsilon_\chi= \epsilon_{\rm kin} + m_\chi n_\chi + \frac{g^2_\chi}{2m^2_\phi}~n^2_\chi \,,
\label{eq:DarkEoS}
\end{equation}
with $\epsilon_{\rm kin}$ the DM kinetic energy. Here we neglect interaction between DM and baryons since we expect $g_B \ll  g_\chi$. For spin $\frac{1}{2}$ dark fermions
\begin{equation}
\epsilon_{\rm kin} = \frac{1}{\pi^2}\int_0^{k_{F\chi}}~k^2 \mathrm{d}k~(\sqrt{k^2+m_\chi^2}-m_\chi) \,,
\label{eq:Echi}
\end{equation}
where the Fermi momentum $k_{F\chi} = (3\pi^2 n_\chi)^{1/3}$.  For bosonic DM, the kinetic energy $\epsilon_{\rm kin}\approx 0$ since bosons occupy the lowest momentum state. The pressure of DM is obtained from the thermodynamic relation $p_\chi = -\epsilon_\chi + \tilde{\mu}_\chi n_\chi$ valid at zero temperature.

Repulsive interactions are necessary to stabilize bosonic DM, while for fermions the degeneracy pressure provides additional stabilization. The Compton wavelength for light mediators with $m_\phi$ in the eV-MeV range is much larger than the inter-particle distance, and the interaction between DM particles is enhanced since they interact with a large number of surrounding particles.

To include the back reaction of DM we adopt the following procedure. First we choose the central DM number density $n_\chi(0)$ and its corresponding local chemical potential $\tilde{\mu}_{\chi} (r=0)$. Using $g_{tt}(r)$ obtained from solving the TOV equations for nuclear matter NSs,  Eq.~\ref{eq:muConstant} the DM density profile is calculated in the local density approximation. The corresponding energy density and pressure contributions due to DM  is obtained and used to update the EOS.  The TOV equation is solved using the updated EOS and repeated iteratively for fixed central baryon number density $n_B(0)$ and $n_\chi(0)$ until convergence.  We find that for total DM mass $M_\chi\lesssim10^{-2}~M_\odot$ effects of DM on the baryon density profile is negligible.

The extended DM halo modifies tidal interactions between merging NSs. Assuming circular orbits during inspiral, the phase shift of GW due to the tidal force is given by
$$\delta\Psi=-\frac{117}{256}v^5\frac{M}{\mu}~\tilde{\Lambda}\,,$$
where
$$\tilde{\Lambda}=\frac{16}{13} \frac{(M_1+12M_2) M_1^4 \Lambda_1+(12M_1+M_2 )M^4_2\Lambda_2}{(M_1+M_2)^5}\,,$$
is the relevant measure of the tidal deformability of the BNS system \cite{Flanagan:2007ix}. Here  $M_{1,2}$ and $\Lambda_{1,2}$ are the masses and dimensionless tidal deformabilities of component stars, $\mu$  is the reduced mass, and $v=(\pi M f)^{1/3}$ the orbital velocity.  The (dimensionful) tidal deformability is often written as $\lambda=\frac{2}{3}k_2~R^5$ where $k_2$ is called the tidal Love number and depends on the internal structure of the object \cite{Love73}.  Although tidal effects change the phase only at the 5th order in post-Newtonian expansion (in which the orbital velocity $v$ is dimensionless expansion parameter),  corrections at this order from point-particle general relativistic dynamics is expected to be significantly smaller for NSs with low spins \cite{Flanagan:2007ix}.

The calculation of $\lambda$ has been discussed extensively and we shall use the main results and notation used in Refs.~\cite{Hinderer:2007mb,Postnikov:2010yn}.  The calculation $k_2$ is  detailed in \cite{Hinderer:2007mb} where it is shown that
\begin{multline}
k_2=\frac{8 C^5}{5} (1 - 2 C)^2 \left(2 (1 - C) + (2 C-1) y_R\right)\times \\
\Bigg\{4 C^3 \left(13 - 11 y_R + 2 C^2 (1 + y_R) + C (-2 + 3 y_R)\right)\\
+2 C \left(6 - 3 y_R + 3 C (5 y_R-8)\right)+3 (1 - 2 C)^2  \\
\times  \left(2 + 2 C (y_R-1) - y_R\right) \ln(1 - 2 C) \Bigg\}^{-1}\,.
\label{eq:k2}
\end{multline}
where $C=M/R$ is the compactness of the NS. $y_R=y(R)$ is related to quadrupolar perturbed metric function and is obtained as a solution to the differential equation
\begin{align}
\frac{dy(r)}{dr}=&-\frac{y(r)^2}{r}-\frac{y(r) g_{rr}(r)}{r}\left(1+4\pi r^2(p(r)-\epsilon(r))\right) \nonumber \\
&-4\pi r  \left(9p(r)+5\epsilon(r)+ \frac{\ud\epsilon}{\ud p}(p(r)+\epsilon(r))\right)g_{rr}(r) \nonumber \\
&+ r \left(\frac{6g_{rr}(r)}{r^2}+ \left(\frac{d \ln g_{tt}(r)}{dr}\right)^2 \right) \,,
\label{eq:yR}
\end{align}
where $g_{rr}(r)$ and $g_{tt}(r)$ defined earlier are the radial and temporal components of the unperturbed Schwarzschild metric. We note that the functions $p(r)$ and $\epsilon(r)$ appearing on the right hand side of Eq.~\ref{eq:yR} are uniquely specified by the hydrostatic structure of the unperturbed star and contain contributions due to baryons and DM as discussed earlier.

\begin{figure}[h]
	\centering
	\includegraphics[width=\linewidth]{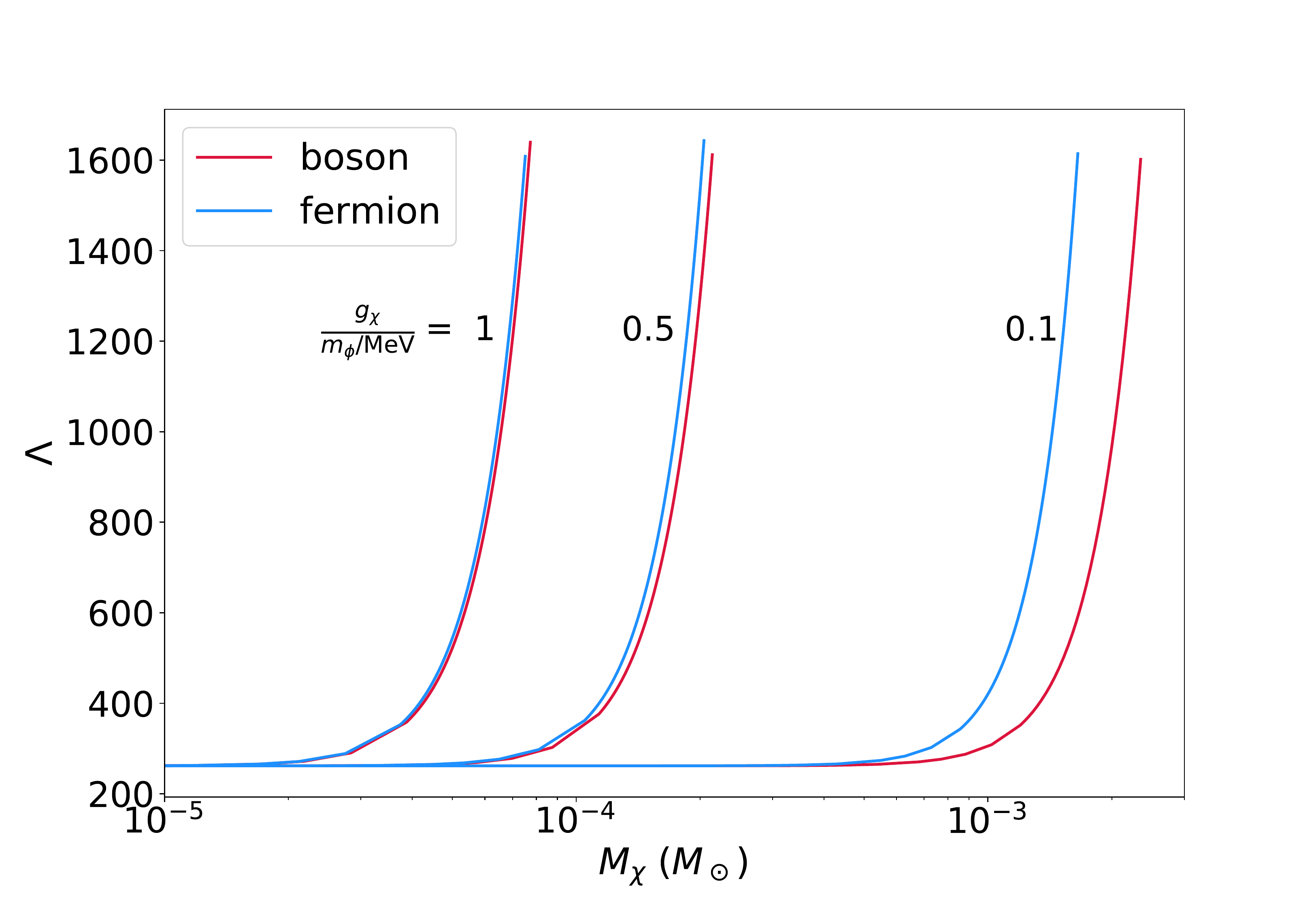}
	\caption{$\Lambda$ increases rapidly with increasing total DM mass $M_\chi$. For self-interacting DM with $g_\chi/m_\phi> 1\  $MeV$^{-1}$,  $M_\chi > 10^{-4} M_\odot$ will boost $\Lambda$ above the upper bound ($\simeq 800$) set by GW170817.}
	\label{fig:lam_M}
\end{figure}

\begin{figure}[h]
	\centering
	\includegraphics[width=\linewidth]{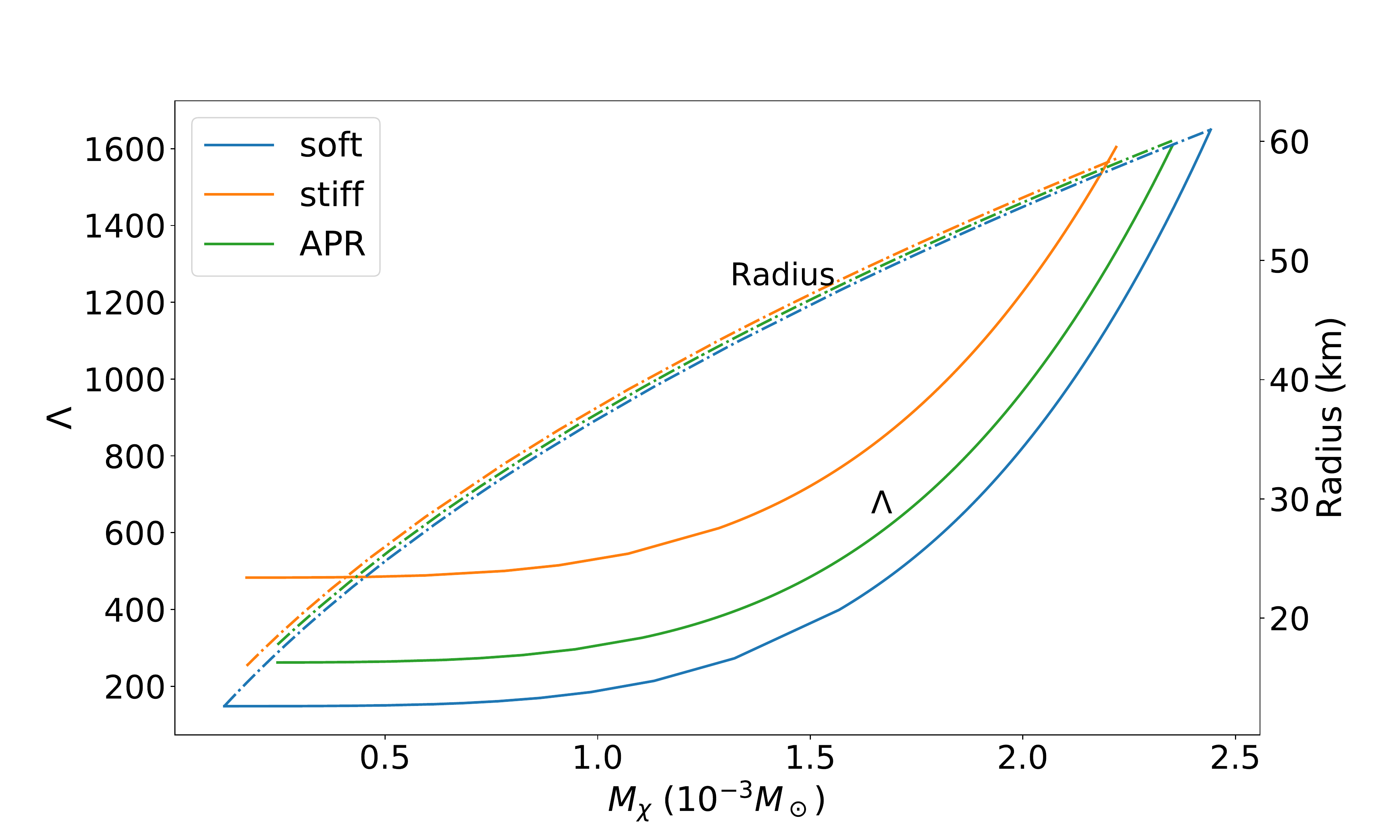}
	\caption{Variation of $\Lambda$ (solid lines) and radius (dashed lines) of a $1.4 M_\odot$ NS for three representative realistic nuclear EOS~\cite{Gandolfi:2011xu} as a function of total dark matter mass $M_\chi$.   vary from $150$ to $500$. dark matter is assumed to be bosonic with $m_\chi=100$ MeV and $g_{\chi}/m_\phi=0.1$ MeV$^{-1}$. The EOS labeled ``soft'' has a smaller pressure at a given density compared to the APR EOS and the EOS labelled ``stiff'' has a higher pressure. The former produces more compact neutron stars with a maximum $\simeq 2~M_\odot$ NS, while the latter predicts an maximum mass  $\simeq 2.4~M_\odot$ and larger radius.}
	\label{fig:eos_dep}
\end{figure}

For reasonable ranges of DM model parameters we find large changes to $\Lambda$ are possible. Results for $m_\chi=100$ MeV and for different interaction strengths are shown in Fig.~\ref{fig:lam_M}. Either strong coupling or light mediator can result in large $\Lambda$ even for trace amounts of DM with total mass $M_\chi \ll M_{NS}$.
The enhancement of $\Lambda$ with increasing total DM mass is remarkable. For small $M_\chi$ the amount of dark matter extending beyond the baryons does not contain much mass and therefore the change in $\Lambda$ is small. However, with increasing $M_\chi$ the dark halo extends further and the mass of dark matter at large radius increases. Since $\lambda = Q_{ij}/ E_{ij}$ where $Q_{ij} \propto \int \eps(r) r^4 dr $ is the quadrupole moment induced by an external tidal $ E_{ij} \propto 1/r^3 $, we expect $\lambda$ to grow rapidly with $M_\chi$ because the halo radius $R_\chi$ increase with $M_\chi$ and the support of energy density $\eps_\chi(r)$ extends further out. Dimensional analysis suggests that $\lambda \propto \int \eps(r) r^7 dr $ and is confirmed by our calculation. As the extent of the dark halo increases, $\lra{\eps_\chi r^7}$ increases rapidly and becomes the dominant contribution to $\lambda$ when it exceed the baryonic contribution $\lra{\eps_B r^7}$. This provides an explanation of the behavior seen in Figs. \ref{fig:lam_M}  and \ref{fig:eos_dep}. Deformations of the halo will then dominate and greatly amplify tidal responses of the hybrid object. This is the main new finding of this study, and it provides a novel probe of the strongly interacting bosonic and fermionic light dark sectors.

Fermion dark halos are bigger and have larger $\Lambda$ due to the additional contribution from the Fermi degeneracy pressure.
For $m_\chi=100$ MeV, the difference between fermions and bosons is modest but the difference increases rapidly with decreasing $m_\chi$. Halos formed by fermions with
$m_\chi\lesssim 30$ MeV can have large $\Lambda$ even without vector mediated self-interactions. For example, $\Lambda_{1.4M_\odot}=800$ is reached for $m_\chi = 30$ MeV  at $M_\chi=10^{-4}M_\odot$, for $m_\chi = 10$ MeV  at $M_\chi=3\times 10^{-6}M_\odot$, and for $m_\chi = 5$ MeV at $M_\chi=4\times10^{-7} M_\odot$. However, the radii of these halos are large. At $\Lambda_{1.4M_\odot}=800$, the halo radius $R\simeq 100$ km for $m_\chi = 30$ MeV, and $R\simeq 210$ km for $m_\chi = 10$ MeV. Since finite size effect during inspiral on waveforms is modeled without accounting for hydrodynamics it is restricted to the phase when the dark halos do not overlap. For this reason we restrict our study to dark halos with radii $R\lesssim150$ km. This will ensure that at lower frequencies detectable by Ad. LIGO the halos do not overlap.

Many-body theories of neutron matter which employ realistic nucleon-nucleon interactions can calculate the EOS up to densities $\simeq 3 \times 10^{14} $ g/cm$^3$ and estimate its error rather reliably\cite{Gandolfi:2011xu,Hebeler:2015hla,Tews:2018kmu}. Using these calculations, and despite large uncertainties associated with the nuclear EOS at higher density, earlier work has shown that the radius of a $1.4~M_\odot$ NS is in the range $10-13$ km \cite{Hebeler:2010jx,Gandolfi:2011xu,Tews:2018kmu}. The corresponding tidal deformability $\Lambda$ is in the range $100-500$. To explore this range we  employ a more general parameterization of neutron matter EOSs based on realistic nucleon-nucleon interactions discussed in \cite{Gandolfi:2011xu} and construct equations of state that are stiffer and softer than the APR EOS discussed earlier. A softer EOS has on average lower pressure for the densities encountered and would result in more compact neutron stars with a smaller maximum mass compared to the APR EOS. In contrast a stiffer EOS will have a larger pressure at a given density compared to the APR EOS and would result in larger NS radii and maximum masses.
The radii and tidal deformabilities of NSs with DM constructed using the three representative nuclear EOSs as a function of $M_\chi$ is shown in Fig.~\ref{fig:eos_dep}. A relatively small increase in the amount of DM allows one to obtain large values of $\Lambda$ inaccessible to ordinary NSs.

\section{Accumulating Dark Matter in a Neutron Star}
\label{sec:acc}
The proto-neutron star formed subsequent to core-collapse during a supernova is characterized by high temperature $T_{\rm PNS}\simeq 30-50$ MeV and density. Under these conditions reactions such as $N N \rightarrow N N  \chi \bar{\chi}$ can produce a large number of dark matter pairs even when the coupling between dark matter and baryons is weak. In our simple model, the interaction between baryons and dark matter particles is mediated by a light vector which couples weakly to baryons  with $g_B \lesssim 10^{-10}$.  The upper bound on $g_B$ is in fact obtained by requiring that the total energy radiated due to the Bremsstrahlung reaction $N N \rightarrow N N \phi$ from the PNS does not exceed $\approx 10^{53}$ ergs  to be compatible with the neutrino signal observed from SN1987 \cite{Rrapaj:2015wgs,Chang:2016ntp,Chang:2018rso}. Since $\phi$ couples strongly to dark matter, pair production through the reaction $N N \rightarrow N N  \chi \bar{\chi}$  will proceed at a rate that is comparable to the Bremsstrahlung reaction if the mass of the
 DM pair is not much larger than $T_{\rm PNS}$.  Thus, a non-negligible fraction of thermal energy of the PNS can be converted into  DM pairs. We estimate that for $g_B = 10^{-10}$ and $m_\chi \lesssim 100\MeV$, the energy radiated as DM pairs is $\approx 10^{51}$ ergs and the corresponding mass of DM produced is $M_\chi= 5.5 \times 10^{-4}~M_\odot$.

In the absence of interactions between dark matter and baryons these pairs will either escape the star when their velocity is larger than the escape velocity $v_{\rm esc}=R_S/R_{\rm NS} \simeq 0.4$ ($R_S=2M_{\rm NS}$ is the Schwarzschild radius of NS) or will annihilate way inside the star. However, the coupling to baryons creates an asymmetry between the energies of dark particles and dark antiparticles and provides a unique mechanism to trap dark charges in NSs.

For the DM model being considered, the energy of $\chi$ and $\bar\chi$ in a dense background of baryons can be calculated using mean field theory. Assuming that dark matter is attracted to baryons and anti-dark matter is repulsed, we find that their energies are given by
\begin{align*}
E_\chi(p)&=\sqrt{p^2+m_\chi^2} - \frac{g_\chi g_B}{m^2_\phi}~n_B \,,\\
E_{\bar{\chi}}(p)&=\sqrt{p^2+m_\chi^2} + \frac{g_\chi g_B}{m^2_\phi}~n_B\,,
\end{align*}
respectively. The energy splitting
\begin{equation}
\begin{aligned}
\Delta E &= E_{\bar{\chi}}- E_\chi= \frac{2g_\chi g_B}{m^2_\phi}~n_B \\
&\simeq 40~\text{MeV} ~\frac{g_\chi g_B}{10^{-11}}\left(\frac{0.1~\text{keV}}{m_\phi}\right)^2 \frac{n_B}{0.5~\text{fm}^{-3}}.
\end{aligned}
\end{equation}
is large and for $m_\phi$ in the range eV - MeV and $g_B g_\chi \simeq 10^{-16}-10^{-10}$ this energy splitting can be comparable to the gravitational binding energy of DM.

This asymmetry will preferentially expel anti-dark matter from the PNS and provide a mechanism to trap a net dark charge in NS that will not be diminished by annihilation. The gradient in the baryon density will induce a force on dark matter towards the core and expel anti-dark matter towards the surface. For example, for model parameters $m_\chi=50$ MeV, $g_\chi=1$, $g_B=10^{-10}$ and $m_\phi = 1$ keV, we find that $80\%$ of $\chi$'s produced during the supernova will remain trapped and survive annihilation while most of the $\bar{\chi}$'s will be accelerated to velocities adequate to escape the PNS.

Once trapped the cloud of DM particles will thermalize among themselves rapidly by exchanging energy and momenta through multiple $\chi \chi \rightarrow \chi \chi$ scatterings. In the born approximation the cross section is
$$\sigma^\text{Born}_{\chi\chi}= \frac{g_\chi^4 m_\chi^2}{4\pi m_\phi^2(m_\phi^2+p^2)}\approx \frac{g_\chi^4 m_\chi}{8\pi m_\phi^2T}. $$
where $p^2\simeq T^2+2m_\chi T\simeq 2m_\chi T$ is the typical momenta of $\chi$'s at temperature $T$, and we assume that $m_\phi^2 \ll m_\chi T$ to obtain the last equality. The corresponding mean free path $\lambda$ is estimated to be
\begin{equation}
\begin{aligned}
\lambda&=\frac{1}{n_\chi\sigma_{\chi\chi}}\simeq 10^{-12}~\text{km} \\
&\times\frac{10^{-4} M_\odot}{M_\chi}
\left(\frac{0.1}{g_\chi} \right)^4
\left(\frac{R_\chi}{10~ \text{km}} \right)^3
\left(\frac{m_\phi}{1 ~\text{keV}} \right)^2
\frac{T}{30~ \text{MeV}}\,,
\end{aligned}
\end{equation}
assuming that the DM density  $n_\chi=(M_\chi/m_\chi) / (4\pi R_\chi^3/3)$. The short mean free path implies that the timescale for thermalization among DM particles $\tau \approx 1/(n_\chi v \sigma_{\chi\chi})$ is much smaller than all other relevant timescales. Further, despite its weaker coupling to baryons the timescale for DM to thermalize with baryons is also very short, typically much less than a second, for reasonable values of the coupling needed to produce adequate DM.  Thus, it is safe to assume that DM in neutron stars shortly after their birth will be characterized by the same low temperature $T < \MeV$  of ordinary matter and any further loss of DM due to evaporation is negligible. Although a detailed study of DM production and charge separation due to interactions with baryons is warranted and will be reported elsewhere, the preceding arguments provide a natural mechanism to generate DM in neutron stars with a mass of the order of $ 10^{-4}~M_\odot$.

Another possibility is that a small fraction of the neutrons in a neutron star decay into dark matter particles which carry baryon number, with the self repulsion of the dark baryons stabilizing the star and allowing for the neutron star to have a mass as large as has been observed \cite{Fornal:2018eol,McKeen:2018xwc,Grinstein:2018ptl}.

Yet another possibility is to link the asymmetry in lepton number or baryon number  in a supernova to the DM asymmetry. To provide an existence proof  for  an  asymmetric production mechanism for ADM   in a supernova,  we constructed the following   simple model. In our model, besides the DM particle $\chi$, which carries dark charge +1, we also introduce a   particle $\Phi$ which carries dark charge $-1$. We also  charge $\chi$ under the global $B-L$ symmetry of the standard model. If $\chi$ is a spin 0 boson then $\Phi$ is a Dirac Fermion, although this scenario will also work if $\chi$ is a Dirac fermion and $\Phi$ is a boson. The dark $U(1)$ will be Higgsed by a field $\xi$ which carries dark charge $2$. In addition, we introduce a   ``right handed" (aka ``sterile'') Dirac  neutrino $N$, which can mix with the standard model neutrinos $\nu$ as in reference \cite{Bertoni:2014mva} and has no standard model gauge interactions.  This mixing will also lead to production of $N$ particles via charged weak currents, with $N$ and $\bar{N}$ being produced asymmetrically due to the lepton asymmetry.  This asymmetry will be converted to an asymmetry in the DM due to the following   interaction term:
\begin{equation}
{\cal L}\supset \lambda N \Phi\chi + g \Phi\Phi\xi + {\rm h.c.}
\end{equation}
These terms are allowed by all symmetries.  The vacuum expectation value of $\xi$ will split the CP even and CP odd components of $\Phi$, leaving a $Z_2$ symmetry unbroken under which  $\chi$ and $\Phi$ are both charged.  We label the two mass eigenstates of $\Phi$ as $\Phi_1, \Phi_2$.  Because $\xi$ is not allowed by B-L number to couple to the DM particle $\chi$, the CP even and odd components of $\chi$ are not split, and $\chi$ continues to carry an exactly conserved $U(1)_D$ global charge.
Provided that  $m_N> m_{\Phi_1}+ m_{\chi}$, the decays $N\rightarrow \bar{\chi} +\Phi_1$ will transmit the asymmetry in the production of $N$ particles to an asymmetry in the production of DM $\chi$ particles.  The $\Phi_1$ particles, which are CP eigenstates,  will decay with equal branching fractions into $\bar\chi+ \nu$ and $\chi+\bar\nu$. Note that a related mechanism in the early universe could lead to genesis of a DM asymmetry from the lepton number asymmetry.

The mechanisms discussed above are generic and would endow all neutron stars with similar amounts of dark matter with perhaps a small variability associated with differences between neutron star and supernova properties. This allows us to constrain the properties of DM using the upper bound on the tidal deformability obtained from GW170817 as we shall discuss  in the next section. However, we note that third generation GW detectors such as the Einstein telescope \cite{Punturo:2010} capable of detecting thousands of BNS mergers within a few years motivate looking for variability in neutron star merger properties. If variability in the tidal deformability is detected in rare events it would imply that some fraction of neutron stars are able to capture large amounts of dark matter. Further, since neutron stars are not estimated to accrete nearly enough dark matter on the average to produce an observable gravitational wave signature if dark matter is distributed uniformly \cite{Kouvaris:2010vv,Kouvaris:2010jy,Kouvaris:2013awa,Cermeno:2017xwb,Kopp:2018jom}, discovering variability would have important implications for small scale structure in dark matter and hint at the existence of compact dark objects.

Such objects are possible in some models of self-interacting DM such as atomic dark matter which contain particles and interactions that can lead to the formation of macroscopic objects through cooling mechanisms analogous to the baryonic sector \cite{Kaplan:2009de}. Constraints on dark compact objects such as clumps of dark matter arise from microlensing surveys. However, even the most stringent constraints derived by the EROS collaboration cannot exclude the possibility that a few percent of dark matter in the galactic halo is in the form of dark compact objects with mass $\gtrsim 10^{-5}~M_\odot$ \cite{Tisserand:2006zx}.  It is conceivably that such compact dark objects can be captured by massive stars or neutron stars, and would provide an explanation should variability in the tidal deformability be seen in future observations of merging neutron stars.

\section{Conclusions} \label{sec:con}

Motivated by the prospect that more BNS and black hole-NS merging events will give better constraints on the tidal deformability $\Lambda$, we have studied how these observations might provide insights about strongly interacting DM.  We have shown that trace amounts of DM with total mass $ M_\chi \gtrsim 10^{-5}~M_\odot$ trapped in a NS, either produced during a supernova, or inherited from a progenitor star, may form a stable dark halo with radii $R\simeq 30-150$ km given sufficient self-repulsion. This can enhance $\Lambda$ of the hybrid to a value significantly larger than expected for ordinary NS, which despite uncertainties associated with the EOS at supra-nuclear density, have $\Lambda_{1.4M_\odot}\lesssim 500$.  Interactions between DM due to exchange of light mediators is coherent and leads to large halos even when $M_\chi$ is small. Further, we find that small variability in the amount of trapped DM can produce large variability in $\Lambda_{1.4M_\odot}$, distinguishing them from ordinary NS where the expected variability is small.  These observable features are unique signatures of strongly interacting DM that can be probed with Advanced LIGO and future GW detectors.

\begin{figure}
	\centering
	\includegraphics[width=\linewidth]{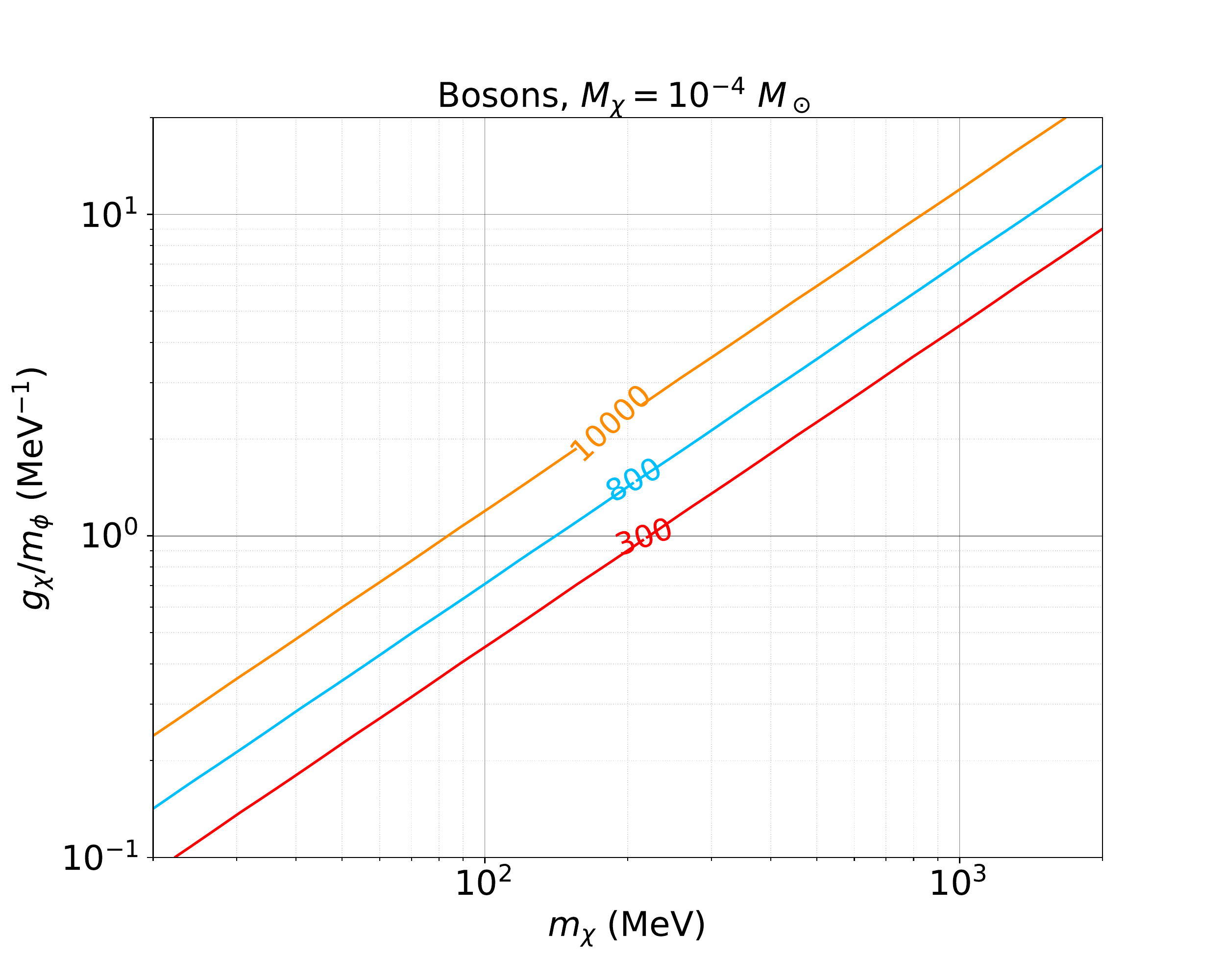}
	\caption{Contours of dimensionless tidal deformability for $1.4~M_\odot$ hybrid stars containing $10^{-4}~M_\odot$ of bosonic DM. The parameter space to the left of the red contour labeled $800$ is disfavored by the first detection of gravitational waves (GW170817) from merging neutron stars.}
	\label{fig:excl_bosons}
\end{figure}

If trace amounts of DM were present in all neutron stars because it was produced during the supernova the constraint from GW170817 that $\Lambda_{1.4M_\odot} < 800$ already excludes a significant fraction of the DM parameter space in our model. The bounds for bosonic DM from our simple model are shown in Fig.~\ref{fig:excl_bosons}.  We have fixed $M_\chi=10^{-4}~M_\odot$ and varied the DM particle mass $m_\chi$ and effective coupling strength $g_\chi/m_\phi$ to obtain the contours of $\Lambda_{1.4M_\odot}$. Bounds for different values of $M_\chi$ can  also be obtained and we found the contour of $\Lambda_{1.4M_\odot}=800$ when $M_\chi\lesssim 0.1M_\odot$ can be approximately fitted by

\begin{equation}
\left(\frac{g_\chi}{m_\phi/\text{MeV}}\right)_{\Lambda_{1.4M_\odot}=800}=1.6\times 10^{-5}\left(\frac{M_\chi}{M_\odot}\right)^{-2/3}\frac{m_\chi}{\text{MeV}} \,.
\end{equation}

\begin{figure}
	\centering
	\includegraphics[width=\linewidth]{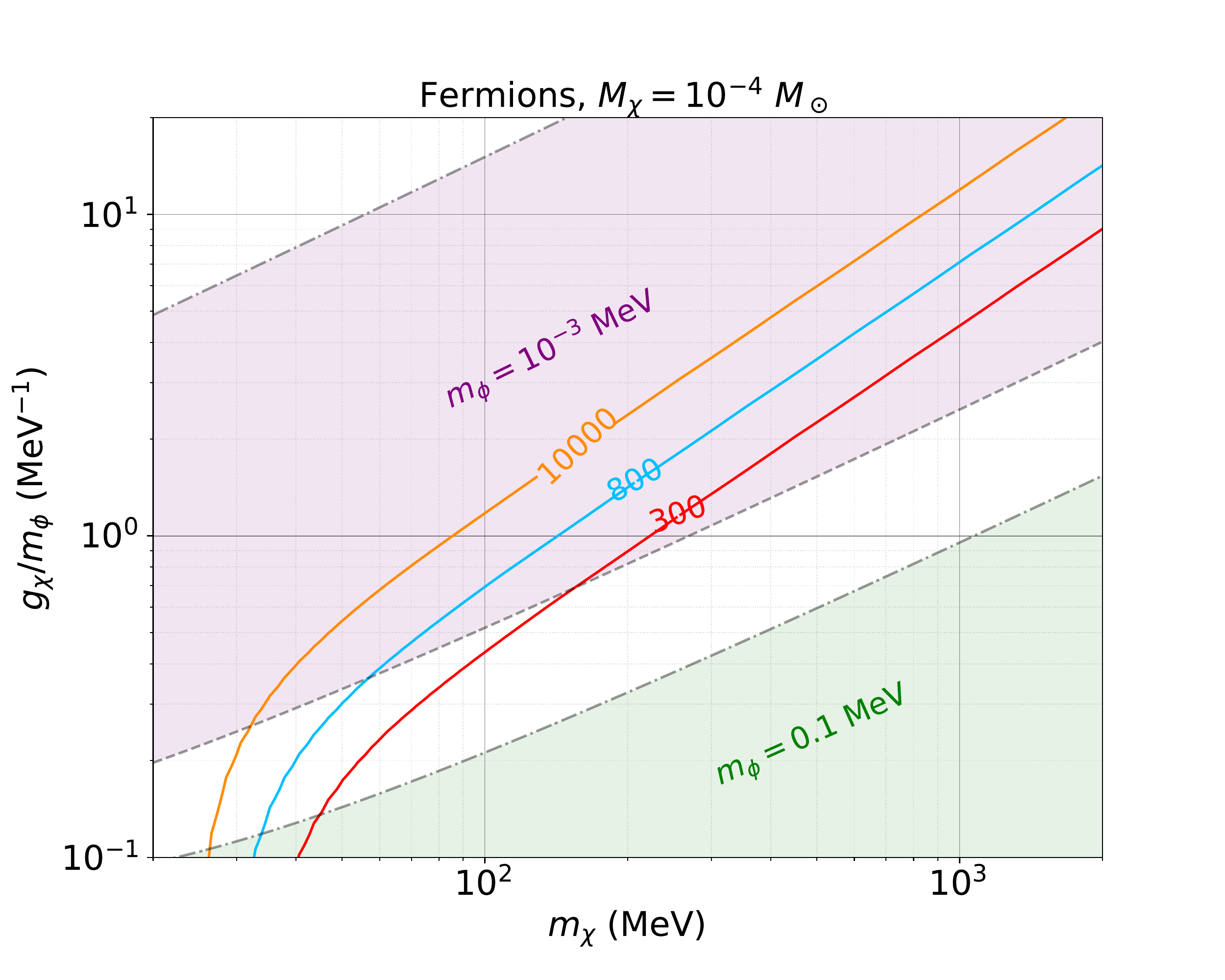}
	\caption{Contours of $\Lambda_{1.4M_\odot}$ for hybrid stars containing $10^{-4}~M_\odot$ of fermionic DM. Regions above dash-dotted lines produce large cross-sections excluded by Bullet Cluster for given mediator masses $m_\phi$, and the shaded regions between the dashed and dash-dotted lines corresponds to the parameter space  where the transport cross section is adequate to provide an explanation of the DM density profiles  in the cores of galaxies and clusters \cite{Kaplinghat:2015aga}.}
	\label{fig:excl_fermions}
\end{figure}

It is remarkable that models with light mediators are severely constrained. For example, a model with $m_\chi=100$ MeV and $m_\phi=1$ eV requires $g_\chi \lesssim 10^{-6}$. These constraints should be especially useful since recent observations of  strong absorption of the Lyman-$\alpha$ radiation from some of the earliest stars corresponding to the 21-centimeter transition of atomic hydrogen around redshift 20 \cite{Bowman:2018fk} appears to favor light DM in the MeV-GeV mass range and whose interactions (with baryons) are due to the exchange of much lighter mediators \cite{Barkana:2018uq}.

Similar bounds can be obtained for fermions as shown in fig.~\ref{fig:excl_fermions}. For heavy fermions, where the contribution due to the Fermi degeneracy pressure is small, contours of $\Lambda$ are very similar to those obtained for bosons. As already noted it is interesting that for light dark fermions with $m_\chi \lesssim 30$ MeV,  $\Lambda_{1.4M_\odot}$ can be larger than 800 even in the absence of strong interactions or light mediators. This is clearly seen in the behavior of contours for small values of $m_\chi$ when they plunge to smaller values of the effective coupling strength $g_\chi/m_\phi$. In all cases we have used the APR EOS to describe the underlying NS. We also set $g_B=0$ in which case the effective Fermi coupling $(g_\chi/m_\phi)^2$ alone determines the interaction strength.

As noted earlier, simulations of large scale structure performed with collision-less DM is unable to account for features on small scales. In particular, simulations predict an excess DM density in the cores of halos of galaxies and clusters, and these discrepancies have been discussed extensively in the literature. For a recent review see \cite{Tulin:2017ara}.  Collisions between DM particles can alleviate some of these issues if the transport cross-section is large enough. For example, a mildly velocity dependent transport cross-section per unit mass, $\sigma_t/m \simeq 2 $ cm$^2$/g on galactic scales, and $\sigma_t/m \simeq 0.1 $ cm$^2$/g on cluster scales has been shown to provide a consistent solution to the dark matter deficit \cite{Kaplinghat:2015aga}. These cross-sections can be consistent with upper limits derived from observations of the Bullet Cluster which requires $\sigma_t/m < 1.25 $ cm$^2$/g  with estimated relative velocity $v_{rel} \approx 3000$ km/s \cite{Randall:2007ph} for velocity dependent interactions. Although in general the tidal deformability provides a complementary constraint, in the scenario where $\chi$ is the sole dark matter particle with Yukawa interactions mediated by a vector boson, as in our simple model, a comparison of these different constraints is useful. To make this comparison we choose two values for the mediator mass: $m_\phi=1$ keV (purple) and $m_\phi=0.1$ MeV (green) in Fig. \ref{fig:excl_fermions}. The shaded parameter region corresponds to values of the transport cross-section that would make N-body simulations of halos of galaxies and clusters compatible with observations. Dashed curves are obtained by setting  $\sigma_t/m \simeq 0.7$ cm$^2$/g for $v_{rel}=4000$ km/s (tighter constraints from Bullet Cluster mass-to-light ratio), and the dashed curve is obtained by setting  $\sigma_t/m \simeq 1 $ cm$^2$/g for $v_{rel}=100$ km/s. The parameter space to the left of the dash-dotted curve violates the cross-section bound from the Bullet Cluster.

With more detections of BNS and black hole-neutron star mergers expected we anticipate that the limits on the tidal deformability will improve and provide stronger constraints. There is also the tantalizing possibility that as the detection sample grows, Ad. LIGO might detect a larger than expected variability in the tidal deformability for neutron stars implicating that some may have dark halos.
We have also proposed mechanisms for DM production and retention that are new. These warrant further study to obtain quantitative estimates for the amount of DM accumulated and its dependence on the model parameters.

Finally, we note that our study only addressed how DM halos influence GW emission through their effects on $\Lambda$ before the DM halos overlap.  The dynamics of larger halos or later times during merger will likely have other interesting observable signatures. We hope our study will motivate the inclusion of DM in numerical relativity simulations to study the complex hydrodynamic evolution of the merger of hybrid stars.

\begin{acknowledgments}
We thank the participants of the INT-JINA symposium on the "First multi-messenger observations of a neutron star merger
and its implications for nuclear physics" for useful discussions and an anonymous referee for alerting us to astrophysical bounds on dark matter cross-sections.  The work of AN was supported by the US Dept. of Energy (DOE) Grant No. DE-SC0011637 and by the Kenneth K. Young chair. SR was supported by DOE Grant No. DE-FG02-00ER41132. DZ was supported by the National Science Foundation Grant No. PHY-1430152 (JINA Center for the Evolution of the Elements).
\end{acknowledgments}
 \bibliographystyle{apsrev4-1}

\begin{thebibliography}{56}%
\makeatletter
\providecommand \@ifxundefined [1]{%
 \@ifx{#1\undefined}
}%
\providecommand \@ifnum [1]{%
 \ifnum #1\expandafter \@firstoftwo
 \else \expandafter \@secondoftwo
 \fi
}%
\providecommand \@ifx [1]{%
 \ifx #1\expandafter \@firstoftwo
 \else \expandafter \@secondoftwo
 \fi
}%
\providecommand \natexlab [1]{#1}%
\providecommand \enquote  [1]{``#1''}%
\providecommand \bibnamefont  [1]{#1}%
\providecommand \bibfnamefont [1]{#1}%
\providecommand \citenamefont [1]{#1}%
\providecommand \href@noop [0]{\@secondoftwo}%
\providecommand \href [0]{\begingroup \@sanitize@url \@href}%
\providecommand \@href[1]{\@@startlink{#1}\@@href}%
\providecommand \@@href[1]{\endgroup#1\@@endlink}%
\providecommand \@sanitize@url [0]{\catcode `\\12\catcode `\$12\catcode
  `\&12\catcode `\#12\catcode `\^12\catcode `\_12\catcode `\%12\relax}%
\providecommand \@@startlink[1]{}%
\providecommand \@@endlink[0]{}%
\providecommand \url  [0]{\begingroup\@sanitize@url \@url }%
\providecommand \@url [1]{\endgroup\@href {#1}{\urlprefix }}%
\providecommand \urlprefix  [0]{URL }%
\providecommand \Eprint [0]{\href }%
\providecommand \doibase [0]{http://dx.doi.org/}%
\providecommand \selectlanguage [0]{\@gobble}%
\providecommand \bibinfo  [0]{\@secondoftwo}%
\providecommand \bibfield  [0]{\@secondoftwo}%
\providecommand \translation [1]{[#1]}%
\providecommand \BibitemOpen [0]{}%
\providecommand \bibitemStop [0]{}%
\providecommand \bibitemNoStop [0]{.\EOS\space}%
\providecommand \EOS [0]{\spacefactor3000\relax}%
\providecommand \BibitemShut  [1]{\csname bibitem#1\endcsname}%
\let\auto@bib@innerbib\@empty
\bibitem [{\citenamefont {Bauer}\ \emph {et~al.}(2015)\citenamefont {Bauer}
  \emph {et~al.}}]{Bauer:2013ihz}%
  \BibitemOpen
  \bibfield  {author} {\bibinfo {author} {\bibfnamefont {D.}~\bibnamefont
  {Bauer}} \emph {et~al.} (\bibinfo {collaboration} {Snowmass 2013 Cosmic
  Frontier Working Groups 1--4}),\ }\href {\doibase 10.1016/j.dark.2015.04.001}
  {\bibfield  {journal} {\bibinfo  {journal} {Phys. Dark Univ.}\ }\textbf
  {\bibinfo {volume} {7-8}},\ \bibinfo {pages} {16} (\bibinfo {year} {2015})},\
  \Eprint {http://arxiv.org/abs/1305.1605} {arXiv:1305.1605 [hep-ph]}
  \BibitemShut {NoStop}%
\bibitem [{\citenamefont {Spergel}\ and\ \citenamefont
  {Steinhardt}(2000)}]{Spergel:2000}%
  \BibitemOpen
  \bibfield  {author} {\bibinfo {author} {\bibfnamefont {D.~N.}\ \bibnamefont
  {Spergel}}\ and\ \bibinfo {author} {\bibfnamefont {P.~J.}\ \bibnamefont
  {Steinhardt}},\ }\href {\doibase 10.1103/PhysRevLett.84.3760} {\bibfield
  {journal} {\bibinfo  {journal} {Phys. Rev. Lett.}\ }\textbf {\bibinfo
  {volume} {84}},\ \bibinfo {pages} {3760} (\bibinfo {year}
  {2000})}\BibitemShut {NoStop}%
\bibitem [{\citenamefont {Loeb}\ and\ \citenamefont
  {Weiner}(2011)}]{Loeb:2011}%
  \BibitemOpen
  \bibfield  {author} {\bibinfo {author} {\bibfnamefont {A.}~\bibnamefont
  {Loeb}}\ and\ \bibinfo {author} {\bibfnamefont {N.}~\bibnamefont {Weiner}},\
  }\href {\doibase 10.1103/PhysRevLett.106.171302} {\bibfield  {journal}
  {\bibinfo  {journal} {Phys. Rev. Lett.}\ }\textbf {\bibinfo {volume} {106}},\
  \bibinfo {pages} {171302} (\bibinfo {year} {2011})}\BibitemShut {NoStop}%
\bibitem [{\citenamefont {Davoudiasl}\ and\ \citenamefont
  {Mohapatra}(2012)}]{Davoudiasl:2012uw}%
  \BibitemOpen
  \bibfield  {author} {\bibinfo {author} {\bibfnamefont {H.}~\bibnamefont
  {Davoudiasl}}\ and\ \bibinfo {author} {\bibfnamefont {R.~N.}\ \bibnamefont
  {Mohapatra}},\ }\href {\doibase 10.1088/1367-2630/14/9/095011} {\bibfield
  {journal} {\bibinfo  {journal} {New J. Phys.}\ }\textbf {\bibinfo {volume}
  {14}},\ \bibinfo {pages} {095011} (\bibinfo {year} {2012})},\ \Eprint
  {http://arxiv.org/abs/1203.1247} {arXiv:1203.1247 [hep-ph]} \BibitemShut
  {NoStop}%
\bibitem [{\citenamefont {Zavala}\ \emph {et~al.}(2013)\citenamefont {Zavala},
  \citenamefont {Vogelsberger},\ and\ \citenamefont {Walker}}]{Zavala_2013}%
  \BibitemOpen
  \bibfield  {author} {\bibinfo {author} {\bibfnamefont {J.}~\bibnamefont
  {Zavala}}, \bibinfo {author} {\bibfnamefont {M.}~\bibnamefont
  {Vogelsberger}}, \ and\ \bibinfo {author} {\bibfnamefont {M.~G.}\
  \bibnamefont {Walker}},\ }\href {\doibase 10.1093/mnrasl/sls053} {\bibfield
  {journal} {\bibinfo  {journal} {Monthly Notices of the Royal Astronomical
  Society: Letters}\ }\textbf {\bibinfo {volume} {431}},\ \bibinfo {pages}
  {L20} (\bibinfo {year} {2013})}\BibitemShut {NoStop}%
\bibitem [{\citenamefont {Bertoni}\ \emph {et~al.}(2015)\citenamefont
  {Bertoni}, \citenamefont {Ipek}, \citenamefont {McKeen},\ and\ \citenamefont
  {Nelson}}]{Bertoni:2014mva}%
  \BibitemOpen
  \bibfield  {author} {\bibinfo {author} {\bibfnamefont {B.}~\bibnamefont
  {Bertoni}}, \bibinfo {author} {\bibfnamefont {S.}~\bibnamefont {Ipek}},
  \bibinfo {author} {\bibfnamefont {D.}~\bibnamefont {McKeen}}, \ and\ \bibinfo
  {author} {\bibfnamefont {A.~E.}\ \bibnamefont {Nelson}},\ }\href {\doibase
  10.1007/JHEP04(2015)170} {\bibfield  {journal} {\bibinfo  {journal} {JHEP}\
  }\textbf {\bibinfo {volume} {04}},\ \bibinfo {pages} {170} (\bibinfo {year}
  {2015})},\ \Eprint {http://arxiv.org/abs/1412.3113} {arXiv:1412.3113
  [hep-ph]} \BibitemShut {NoStop}%
\bibitem [{\citenamefont {Petraki}\ \emph {et~al.}(2014)\citenamefont
  {Petraki}, \citenamefont {Pearce},\ and\ \citenamefont
  {Kusenko}}]{Petraki:2014uza}%
  \BibitemOpen
  \bibfield  {author} {\bibinfo {author} {\bibfnamefont {K.}~\bibnamefont
  {Petraki}}, \bibinfo {author} {\bibfnamefont {L.}~\bibnamefont {Pearce}}, \
  and\ \bibinfo {author} {\bibfnamefont {A.}~\bibnamefont {Kusenko}},\ }\href
  {\doibase 10.1088/1475-7516/2014/07/039} {\bibfield  {journal} {\bibinfo
  {journal} {JCAP}\ }\textbf {\bibinfo {volume} {1407}},\ \bibinfo {pages}
  {039} (\bibinfo {year} {2014})},\ \Eprint {http://arxiv.org/abs/1403.1077}
  {arXiv:1403.1077 [hep-ph]} \BibitemShut {NoStop}%
\bibitem [{Buc(2014)}]{Buckley:2014hja}%
  \BibitemOpen
  \href {\doibase 10.1103/PhysRevD.90.043524} {\bibfield  {journal} {\bibinfo
  {journal} {Phys. Rev.}\ }\textbf {\bibinfo {volume} {D90}},\ \bibinfo {pages}
  {043524} (\bibinfo {year} {2014})},\ \Eprint {http://arxiv.org/abs/1405.2075}
  {arXiv:1405.2075 [astro-ph.CO]} \BibitemShut {NoStop}%
\bibitem [{\citenamefont {Creasey}\ \emph {et~al.}(2017)\citenamefont
  {Creasey}, \citenamefont {Sameie}, \citenamefont {Sales}, \citenamefont {Yu},
  \citenamefont {Vogelsberger},\ and\ \citenamefont
  {Zavala}}]{Creasey:2016jaq}%
  \BibitemOpen
  \bibfield  {author} {\bibinfo {author} {\bibfnamefont {P.}~\bibnamefont
  {Creasey}}, \bibinfo {author} {\bibfnamefont {O.}~\bibnamefont {Sameie}},
  \bibinfo {author} {\bibfnamefont {L.~V.}\ \bibnamefont {Sales}}, \bibinfo
  {author} {\bibfnamefont {H.-B.}\ \bibnamefont {Yu}}, \bibinfo {author}
  {\bibfnamefont {M.}~\bibnamefont {Vogelsberger}}, \ and\ \bibinfo {author}
  {\bibfnamefont {J.}~\bibnamefont {Zavala}},\ }\href {\doibase
  10.1093/mnras/stx522} {\bibfield  {journal} {\bibinfo  {journal} {Mon. Not.
  Roy. Astron. Soc.}\ }\textbf {\bibinfo {volume} {468}},\ \bibinfo {pages}
  {2283} (\bibinfo {year} {2017})},\ \Eprint {http://arxiv.org/abs/1612.03903}
  {arXiv:1612.03903 [astro-ph.GA]} \BibitemShut {NoStop}%
\bibitem [{\citenamefont {Di~Cintio}\ \emph {et~al.}(2017)\citenamefont
  {Di~Cintio}, \citenamefont {Tremmel}, \citenamefont {Governato},
  \citenamefont {Pontzen}, \citenamefont {Zavala}, \citenamefont
  {Bastidas~Fry}, \citenamefont {Brooks},\ and\ \citenamefont
  {Vogelsberger}}]{DiCintio:2017zdz}%
  \BibitemOpen
  \bibfield  {author} {\bibinfo {author} {\bibfnamefont {A.}~\bibnamefont
  {Di~Cintio}}, \bibinfo {author} {\bibfnamefont {M.}~\bibnamefont {Tremmel}},
  \bibinfo {author} {\bibfnamefont {F.}~\bibnamefont {Governato}}, \bibinfo
  {author} {\bibfnamefont {A.}~\bibnamefont {Pontzen}}, \bibinfo {author}
  {\bibfnamefont {J.}~\bibnamefont {Zavala}}, \bibinfo {author} {\bibfnamefont
  {A.}~\bibnamefont {Bastidas~Fry}}, \bibinfo {author} {\bibfnamefont
  {A.}~\bibnamefont {Brooks}}, \ and\ \bibinfo {author} {\bibfnamefont
  {M.}~\bibnamefont {Vogelsberger}},\ }\href {\doibase 10.1093/mnras/stx1043}
  {\bibfield  {journal} {\bibinfo  {journal} {Mon. Not. Roy. Astron. Soc.}\
  }\textbf {\bibinfo {volume} {469}},\ \bibinfo {pages} {2845} (\bibinfo {year}
  {2017})},\ \Eprint {http://arxiv.org/abs/1701.04410} {arXiv:1701.04410
  [astro-ph.GA]} \BibitemShut {NoStop}%
\bibitem [{\citenamefont {Tulin}\ and\ \citenamefont
  {Yu}(2018)}]{Tulin:2017ara}%
  \BibitemOpen
  \bibfield  {author} {\bibinfo {author} {\bibfnamefont {S.}~\bibnamefont
  {Tulin}}\ and\ \bibinfo {author} {\bibfnamefont {H.-B.}\ \bibnamefont {Yu}},\
  }\href {\doibase 10.1016/j.physrep.2017.11.004} {\bibfield  {journal}
  {\bibinfo  {journal} {Phys. Rept.}\ }\textbf {\bibinfo {volume} {730}},\
  \bibinfo {pages} {1} (\bibinfo {year} {2018})},\ \Eprint
  {http://arxiv.org/abs/1705.02358} {arXiv:1705.02358 [hep-ph]} \BibitemShut
  {NoStop}%
\bibitem [{\citenamefont {Duerr}\ \emph {et~al.}(2018)\citenamefont {Duerr},
  \citenamefont {Schmidt-Hoberg},\ and\ \citenamefont {Wild}}]{Duerr:2018mbd}%
  \BibitemOpen
  \bibfield  {author} {\bibinfo {author} {\bibfnamefont {M.}~\bibnamefont
  {Duerr}}, \bibinfo {author} {\bibfnamefont {K.}~\bibnamefont
  {Schmidt-Hoberg}}, \ and\ \bibinfo {author} {\bibfnamefont {S.}~\bibnamefont
  {Wild}},\ }\href@noop {} {\enquote {\bibinfo {title} {{Self-interacting dark
  matter with a stable vector mediator}},}\ } (\bibinfo {year} {2018}),\
  \bibinfo {note} {unpublished},\ \Eprint {http://arxiv.org/abs/1804.10385}
  {arXiv:1804.10385 [hep-ph]} \BibitemShut {NoStop}%
\bibitem [{\citenamefont {Kaplinghat}\ \emph {et~al.}(2016)\citenamefont
  {Kaplinghat}, \citenamefont {Tulin},\ and\ \citenamefont
  {Yu}}]{Kaplinghat:2015aga}%
  \BibitemOpen
  \bibfield  {author} {\bibinfo {author} {\bibfnamefont {M.}~\bibnamefont
  {Kaplinghat}}, \bibinfo {author} {\bibfnamefont {S.}~\bibnamefont {Tulin}}, \
  and\ \bibinfo {author} {\bibfnamefont {H.-B.}\ \bibnamefont {Yu}},\ }\href
  {\doibase 10.1103/PhysRevLett.116.041302} {\bibfield  {journal} {\bibinfo
  {journal} {Phys. Rev. Lett.}\ }\textbf {\bibinfo {volume} {116}},\ \bibinfo
  {pages} {041302} (\bibinfo {year} {2016})},\ \Eprint
  {http://arxiv.org/abs/1508.03339} {arXiv:1508.03339 [astro-ph.CO]}
  \BibitemShut {NoStop}%
\bibitem [{\citenamefont {Kaplan}\ \emph {et~al.}(2009)\citenamefont {Kaplan},
  \citenamefont {Luty},\ and\ \citenamefont {Zurek}}]{Kaplan_2009}%
  \BibitemOpen
  \bibfield  {author} {\bibinfo {author} {\bibfnamefont {D.~E.}\ \bibnamefont
  {Kaplan}}, \bibinfo {author} {\bibfnamefont {M.~A.}\ \bibnamefont {Luty}}, \
  and\ \bibinfo {author} {\bibfnamefont {K.~M.}\ \bibnamefont {Zurek}},\ }\href
  {\doibase 10.1103/physrevd.79.115016} {\bibfield  {journal} {\bibinfo
  {journal} {Physical Review D}\ }\textbf {\bibinfo {volume} {79}} (\bibinfo
  {year} {2009}),\ 10.1103/physrevd.79.115016}\BibitemShut {NoStop}%
\bibitem [{\citenamefont {Boucenna}\ and\ \citenamefont
  {Morisi}(2014)}]{Boucenna:2013wba}%
  \BibitemOpen
  \bibfield  {author} {\bibinfo {author} {\bibfnamefont {S.~M.}\ \bibnamefont
  {Boucenna}}\ and\ \bibinfo {author} {\bibfnamefont {S.}~\bibnamefont
  {Morisi}},\ }\href {\doibase 10.3389/fphy.2013.00033} {\bibfield  {journal}
  {\bibinfo  {journal} {Front.in Phys.}\ }\textbf {\bibinfo {volume} {1}},\
  \bibinfo {pages} {33} (\bibinfo {year} {2014})},\ \Eprint
  {http://arxiv.org/abs/1310.1904} {arXiv:1310.1904 [hep-ph]} \BibitemShut
  {NoStop}%
\bibitem [{\citenamefont {Zurek}(2014)}]{Zurek:2014}%
  \BibitemOpen
  \bibfield  {author} {\bibinfo {author} {\bibfnamefont {K.~M.}\ \bibnamefont
  {Zurek}},\ }\href {\doibase https://doi.org/10.1016/j.physrep.2013.12.001}
  {\bibfield  {journal} {\bibinfo  {journal} {Physics Reports}\ }\textbf
  {\bibinfo {volume} {537}},\ \bibinfo {pages} {91 } (\bibinfo {year}
  {2014})},\ \bibinfo {note} {asymmetric Dark Matter: Theories, signatures, and
  constraints}\BibitemShut {NoStop}%
\bibitem [{\citenamefont {Kouvaris}\ and\ \citenamefont
  {Tinyakov}(2010{\natexlab{a}})}]{Kouvaris:2010}%
  \BibitemOpen
  \bibfield  {author} {\bibinfo {author} {\bibfnamefont {C.}~\bibnamefont
  {Kouvaris}}\ and\ \bibinfo {author} {\bibfnamefont {P.}~\bibnamefont
  {Tinyakov}},\ }\href {\doibase 10.1103/PhysRevD.82.063531} {\bibfield
  {journal} {\bibinfo  {journal} {Phys. Rev. D}\ }\textbf {\bibinfo {volume}
  {82}},\ \bibinfo {pages} {063531} (\bibinfo {year}
  {2010}{\natexlab{a}})}\BibitemShut {NoStop}%
\bibitem [{\citenamefont {Leung}\ \emph {et~al.}(2011)\citenamefont {Leung},
  \citenamefont {Chu},\ and\ \citenamefont {Lin}}]{Leung_2011}%
  \BibitemOpen
  \bibfield  {author} {\bibinfo {author} {\bibfnamefont {S.-C.}\ \bibnamefont
  {Leung}}, \bibinfo {author} {\bibfnamefont {M.-C.}\ \bibnamefont {Chu}}, \
  and\ \bibinfo {author} {\bibfnamefont {L.-M.}\ \bibnamefont {Lin}},\ }\href
  {\doibase 10.1103/physrevd.84.107301} {\bibfield  {journal} {\bibinfo
  {journal} {Physical Review D}\ }\textbf {\bibinfo {volume} {84}} (\bibinfo
  {year} {2011}),\ 10.1103/physrevd.84.107301}\BibitemShut {NoStop}%
\bibitem [{\citenamefont {Kouvaris}\ and\ \citenamefont
  {Nielsen}(2015)}]{Kouvaris_2015}%
  \BibitemOpen
  \bibfield  {author} {\bibinfo {author} {\bibfnamefont {C.}~\bibnamefont
  {Kouvaris}}\ and\ \bibinfo {author} {\bibfnamefont {N.~G.}\ \bibnamefont
  {Nielsen}},\ }\href {\doibase 10.1103/physrevd.92.063526} {\bibfield
  {journal} {\bibinfo  {journal} {Physical Review D}\ }\textbf {\bibinfo
  {volume} {92}} (\bibinfo {year} {2015}),\
  10.1103/physrevd.92.063526}\BibitemShut {NoStop}%
\bibitem [{\citenamefont {Bertoni}\ \emph {et~al.}(2013)\citenamefont
  {Bertoni}, \citenamefont {Nelson},\ and\ \citenamefont
  {Reddy}}]{Bertoni:2013bsa}%
  \BibitemOpen
  \bibfield  {author} {\bibinfo {author} {\bibfnamefont {B.}~\bibnamefont
  {Bertoni}}, \bibinfo {author} {\bibfnamefont {A.~E.}\ \bibnamefont {Nelson}},
  \ and\ \bibinfo {author} {\bibfnamefont {S.}~\bibnamefont {Reddy}},\ }\href
  {\doibase 10.1103/PhysRevD.88.123505} {\bibfield  {journal} {\bibinfo
  {journal} {Phys. Rev.}\ }\textbf {\bibinfo {volume} {D88}},\ \bibinfo {pages}
  {123505} (\bibinfo {year} {2013})},\ \Eprint {http://arxiv.org/abs/1309.1721}
  {arXiv:1309.1721 [hep-ph]} \BibitemShut {NoStop}%
\bibitem [{\citenamefont {Goldman}\ and\ \citenamefont
  {Nussinov}(1989)}]{Goldman:1989nd}%
  \BibitemOpen
  \bibfield  {author} {\bibinfo {author} {\bibfnamefont {I.}~\bibnamefont
  {Goldman}}\ and\ \bibinfo {author} {\bibfnamefont {S.}~\bibnamefont
  {Nussinov}},\ }\href {\doibase 10.1103/PhysRevD.40.3221} {\bibfield
  {journal} {\bibinfo  {journal} {Phys. Rev.}\ }\textbf {\bibinfo {volume}
  {D40}},\ \bibinfo {pages} {3221} (\bibinfo {year} {1989})}\BibitemShut
  {NoStop}%
\bibitem [{\citenamefont {Gould}\ \emph {et~al.}(1990)\citenamefont {Gould},
  \citenamefont {Draine}, \citenamefont {Romani},\ and\ \citenamefont
  {Nussinov}}]{Gould:1989gw}%
  \BibitemOpen
  \bibfield  {author} {\bibinfo {author} {\bibfnamefont {A.}~\bibnamefont
  {Gould}}, \bibinfo {author} {\bibfnamefont {B.~T.}\ \bibnamefont {Draine}},
  \bibinfo {author} {\bibfnamefont {R.~W.}\ \bibnamefont {Romani}}, \ and\
  \bibinfo {author} {\bibfnamefont {S.}~\bibnamefont {Nussinov}},\ }\href
  {\doibase 10.1016/0370-2693(90)91745-W} {\bibfield  {journal} {\bibinfo
  {journal} {Phys. Lett.}\ }\textbf {\bibinfo {volume} {B238}},\ \bibinfo
  {pages} {337} (\bibinfo {year} {1990})}\BibitemShut {NoStop}%
\bibitem [{\citenamefont {Jamison}(2013)}]{Jamison:2013yya}%
  \BibitemOpen
  \bibfield  {author} {\bibinfo {author} {\bibfnamefont {A.~O.}\ \bibnamefont
  {Jamison}},\ }\href {\doibase 10.1103/PhysRevD.88.035004} {\bibfield
  {journal} {\bibinfo  {journal} {Phys. Rev.}\ }\textbf {\bibinfo {volume}
  {D88}},\ \bibinfo {pages} {035004} (\bibinfo {year} {2013})},\ \Eprint
  {http://arxiv.org/abs/1304.3773} {arXiv:1304.3773 [hep-ph]} \BibitemShut
  {NoStop}%
\bibitem [{\citenamefont {McDermott}\ \emph {et~al.}(2012)\citenamefont
  {McDermott}, \citenamefont {Yu},\ and\ \citenamefont
  {Zurek}}]{McDermott:2011jp}%
  \BibitemOpen
  \bibfield  {author} {\bibinfo {author} {\bibfnamefont {S.~D.}\ \bibnamefont
  {McDermott}}, \bibinfo {author} {\bibfnamefont {H.-B.}\ \bibnamefont {Yu}}, \
  and\ \bibinfo {author} {\bibfnamefont {K.~M.}\ \bibnamefont {Zurek}},\ }\href
  {\doibase 10.1103/PhysRevD.85.023519} {\bibfield  {journal} {\bibinfo
  {journal} {Phys. Rev.}\ }\textbf {\bibinfo {volume} {D85}},\ \bibinfo {pages}
  {023519} (\bibinfo {year} {2012})},\ \Eprint {http://arxiv.org/abs/1103.5472}
  {arXiv:1103.5472 [hep-ph]} \BibitemShut {NoStop}%
\bibitem [{\citenamefont {Bell}\ \emph {et~al.}(2013)\citenamefont {Bell},
  \citenamefont {Melatos},\ and\ \citenamefont {Petraki}}]{Bell:2013xk}%
  \BibitemOpen
  \bibfield  {author} {\bibinfo {author} {\bibfnamefont {N.~F.}\ \bibnamefont
  {Bell}}, \bibinfo {author} {\bibfnamefont {A.}~\bibnamefont {Melatos}}, \
  and\ \bibinfo {author} {\bibfnamefont {K.}~\bibnamefont {Petraki}},\ }\href
  {\doibase 10.1103/PhysRevD.87.123507} {\bibfield  {journal} {\bibinfo
  {journal} {Phys. Rev.}\ }\textbf {\bibinfo {volume} {D87}},\ \bibinfo {pages}
  {123507} (\bibinfo {year} {2013})},\ \Eprint {http://arxiv.org/abs/1301.6811}
  {arXiv:1301.6811 [hep-ph]} \BibitemShut {NoStop}%
\bibitem [{\citenamefont {Rrapaj}\ and\ \citenamefont
  {Reddy}(2016)}]{Rrapaj:2015wgs}%
  \BibitemOpen
  \bibfield  {author} {\bibinfo {author} {\bibfnamefont {E.}~\bibnamefont
  {Rrapaj}}\ and\ \bibinfo {author} {\bibfnamefont {S.}~\bibnamefont {Reddy}},\
  }\href {\doibase 10.1103/PhysRevC.94.045805} {\bibfield  {journal} {\bibinfo
  {journal} {Phys. Rev.}\ }\textbf {\bibinfo {volume} {C94}},\ \bibinfo {pages}
  {045805} (\bibinfo {year} {2016})},\ \Eprint
  {http://arxiv.org/abs/1511.09136} {arXiv:1511.09136 [nucl-th]} \BibitemShut
  {NoStop}%
\bibitem [{\citenamefont {Ellis}\ \emph {et~al.}(2018)\citenamefont {Ellis},
  \citenamefont {Hektor}, \citenamefont {H{\"u}tsi}, \citenamefont {Kannike},
  \citenamefont {Marzola}, \citenamefont {Raidal},\ and\ \citenamefont
  {Vaskonen}}]{Ellis:2017jgp}%
  \BibitemOpen
  \bibfield  {author} {\bibinfo {author} {\bibfnamefont {J.}~\bibnamefont
  {Ellis}}, \bibinfo {author} {\bibfnamefont {A.}~\bibnamefont {Hektor}},
  \bibinfo {author} {\bibfnamefont {G.}~\bibnamefont {H{\"u}tsi}}, \bibinfo
  {author} {\bibfnamefont {K.}~\bibnamefont {Kannike}}, \bibinfo {author}
  {\bibfnamefont {L.}~\bibnamefont {Marzola}}, \bibinfo {author} {\bibfnamefont
  {M.}~\bibnamefont {Raidal}}, \ and\ \bibinfo {author} {\bibfnamefont
  {V.}~\bibnamefont {Vaskonen}},\ }\href {\doibase
  10.1016/j.physletb.2018.04.048} {\bibfield  {journal} {\bibinfo  {journal}
  {Phys. Lett.}\ }\textbf {\bibinfo {volume} {B781}},\ \bibinfo {pages} {607}
  (\bibinfo {year} {2018})},\ \Eprint {http://arxiv.org/abs/1710.05540}
  {arXiv:1710.05540 [astro-ph.CO]} \BibitemShut {NoStop}%
\bibitem [{\citenamefont {Flanagan}\ and\ \citenamefont
  {Hinderer}(2008)}]{Flanagan:2007ix}%
  \BibitemOpen
  \bibfield  {author} {\bibinfo {author} {\bibfnamefont {E.~E.}\ \bibnamefont
  {Flanagan}}\ and\ \bibinfo {author} {\bibfnamefont {T.}~\bibnamefont
  {Hinderer}},\ }\href {\doibase 10.1103/PhysRevD.77.021502} {\bibfield
  {journal} {\bibinfo  {journal} {Phys. Rev.}\ }\textbf {\bibinfo {volume}
  {D77}},\ \bibinfo {pages} {021502} (\bibinfo {year} {2008})},\ \Eprint
  {http://arxiv.org/abs/0709.1915} {arXiv:0709.1915 [astro-ph]} \BibitemShut
  {NoStop}%
\bibitem [{\citenamefont {Abbott}\ \emph {et~al.}(2017)\citenamefont {Abbott}
  \emph {et~al.}}]{TheLIGOScientific:2017qsa}%
  \BibitemOpen
  \bibfield  {author} {\bibinfo {author} {\bibfnamefont {B.}~\bibnamefont
  {Abbott}} \emph {et~al.} (\bibinfo {collaboration} {Virgo, LIGO
  Scientific}),\ }\href {\doibase 10.1103/PhysRevLett.119.161101} {\bibfield
  {journal} {\bibinfo  {journal} {Phys. Rev. Lett.}\ }\textbf {\bibinfo
  {volume} {119}},\ \bibinfo {pages} {161101} (\bibinfo {year} {2017})},\
  \Eprint {http://arxiv.org/abs/1710.05832} {arXiv:1710.05832 [gr-qc]}
  \BibitemShut {NoStop}%
\bibitem [{\citenamefont {Abbott}\ \emph {et~al.}(2018)\citenamefont {Abbott}
  \emph {et~al.}}]{Abbott:2018exr}%
  \BibitemOpen
  \bibfield  {author} {\bibinfo {author} {\bibfnamefont {B.~P.}\ \bibnamefont
  {Abbott}} \emph {et~al.} (\bibinfo {collaboration} {LIGO Scientific,
  Virgo}),\ }\href {\doibase 10.1103/PhysRevLett.121.161101} {\bibfield
  {journal} {\bibinfo  {journal} {Phys. Rev. Lett.}\ }\textbf {\bibinfo
  {volume} {121}},\ \bibinfo {pages} {161101} (\bibinfo {year} {2018})},\
  \Eprint {http://arxiv.org/abs/1805.11581} {arXiv:1805.11581 [gr-qc]}
  \BibitemShut {NoStop}%
\bibitem [{\citenamefont {Tolman}(1939)}]{Tolman:1939jz}%
  \BibitemOpen
  \bibfield  {author} {\bibinfo {author} {\bibfnamefont {R.~C.}\ \bibnamefont
  {Tolman}},\ }\href {\doibase 10.1103/PhysRev.55.364} {\bibfield  {journal}
  {\bibinfo  {journal} {Phys. Rev.}\ }\textbf {\bibinfo {volume} {55}},\
  \bibinfo {pages} {364} (\bibinfo {year} {1939})}\BibitemShut {NoStop}%
\bibitem [{\citenamefont {Oppenheimer}\ and\ \citenamefont
  {Volkoff}(1939)}]{Oppenheimer:1939ne}%
  \BibitemOpen
  \bibfield  {author} {\bibinfo {author} {\bibfnamefont {J.~R.}\ \bibnamefont
  {Oppenheimer}}\ and\ \bibinfo {author} {\bibfnamefont {G.~M.}\ \bibnamefont
  {Volkoff}},\ }\href {\doibase 10.1103/PhysRev.55.374} {\bibfield  {journal}
  {\bibinfo  {journal} {Phys. Rev.}\ }\textbf {\bibinfo {volume} {55}},\
  \bibinfo {pages} {374} (\bibinfo {year} {1939})}\BibitemShut {NoStop}%
\bibitem [{\citenamefont {Akmal}\ \emph {et~al.}(1998)\citenamefont {Akmal},
  \citenamefont {Pandharipande},\ and\ \citenamefont
  {Ravenhall}}]{Akmal:1998cf}%
  \BibitemOpen
  \bibfield  {author} {\bibinfo {author} {\bibfnamefont {A.}~\bibnamefont
  {Akmal}}, \bibinfo {author} {\bibfnamefont {V.~R.}\ \bibnamefont
  {Pandharipande}}, \ and\ \bibinfo {author} {\bibfnamefont {D.~G.}\
  \bibnamefont {Ravenhall}},\ }\href {\doibase 10.1103/PhysRevC.58.1804}
  {\bibfield  {journal} {\bibinfo  {journal} {Phys. Rev.}\ }\textbf {\bibinfo
  {volume} {C58}},\ \bibinfo {pages} {1804} (\bibinfo {year} {1998})},\ \Eprint
  {http://arxiv.org/abs/nucl-th/9804027} {arXiv:nucl-th/9804027 [nucl-th]}
  \BibitemShut {NoStop}%
\bibitem [{\citenamefont {Love}(1909)}]{Love73}%
  \BibitemOpen
  \bibfield  {author} {\bibinfo {author} {\bibnamefont {Love}},\ }\href
  {\doibase 10.1098/rspa.1909.0008} {\bibfield  {journal} {\bibinfo  {journal}
  {Proceedings of the Royal Society of London A: Mathematical, Physical and
  Engineering Sciences}\ }\textbf {\bibinfo {volume} {82}},\ \bibinfo {pages}
  {73} (\bibinfo {year} {1909})}\BibitemShut {NoStop}%
\bibitem [{\citenamefont {Hinderer}(2008)}]{Hinderer:2007mb}%
  \BibitemOpen
  \bibfield  {author} {\bibinfo {author} {\bibfnamefont {T.}~\bibnamefont
  {Hinderer}},\ }\href {\doibase 10.1086/533487} {\bibfield  {journal}
  {\bibinfo  {journal} {Astrophys. J.}\ }\textbf {\bibinfo {volume} {677}},\
  \bibinfo {pages} {1216} (\bibinfo {year} {2008})},\ \Eprint
  {http://arxiv.org/abs/0711.2420} {arXiv:0711.2420 [astro-ph]} \BibitemShut
  {NoStop}%
\bibitem [{\citenamefont {Postnikov}\ \emph {et~al.}(2010)\citenamefont
  {Postnikov}, \citenamefont {Prakash},\ and\ \citenamefont
  {Lattimer}}]{Postnikov:2010yn}%
  \BibitemOpen
  \bibfield  {author} {\bibinfo {author} {\bibfnamefont {S.}~\bibnamefont
  {Postnikov}}, \bibinfo {author} {\bibfnamefont {M.}~\bibnamefont {Prakash}},
  \ and\ \bibinfo {author} {\bibfnamefont {J.~M.}\ \bibnamefont {Lattimer}},\
  }\href {\doibase 10.1103/PhysRevD.82.024016} {\bibfield  {journal} {\bibinfo
  {journal} {Phys. Rev.}\ }\textbf {\bibinfo {volume} {D82}},\ \bibinfo {pages}
  {024016} (\bibinfo {year} {2010})},\ \Eprint {http://arxiv.org/abs/1004.5098}
  {arXiv:1004.5098 [astro-ph.SR]} \BibitemShut {NoStop}%
\bibitem [{\citenamefont {Gandolfi}\ \emph {et~al.}(2012)\citenamefont
  {Gandolfi}, \citenamefont {Carlson},\ and\ \citenamefont
  {Reddy}}]{Gandolfi:2011xu}%
  \BibitemOpen
  \bibfield  {author} {\bibinfo {author} {\bibfnamefont {S.}~\bibnamefont
  {Gandolfi}}, \bibinfo {author} {\bibfnamefont {J.}~\bibnamefont {Carlson}}, \
  and\ \bibinfo {author} {\bibfnamefont {S.}~\bibnamefont {Reddy}},\ }\href
  {\doibase 10.1103/PhysRevC.85.032801} {\bibfield  {journal} {\bibinfo
  {journal} {Phys. Rev.}\ }\textbf {\bibinfo {volume} {C85}},\ \bibinfo {pages}
  {032801} (\bibinfo {year} {2012})},\ \Eprint {http://arxiv.org/abs/1101.1921}
  {arXiv:1101.1921 [nucl-th]} \BibitemShut {NoStop}%
\bibitem [{\citenamefont {Hebeler}\ \emph {et~al.}(2015)\citenamefont
  {Hebeler}, \citenamefont {Holt}, \citenamefont {Menendez},\ and\
  \citenamefont {Schwenk}}]{Hebeler:2015hla}%
  \BibitemOpen
  \bibfield  {author} {\bibinfo {author} {\bibfnamefont {K.}~\bibnamefont
  {Hebeler}}, \bibinfo {author} {\bibfnamefont {J.~D.}\ \bibnamefont {Holt}},
  \bibinfo {author} {\bibfnamefont {J.}~\bibnamefont {Menendez}}, \ and\
  \bibinfo {author} {\bibfnamefont {A.}~\bibnamefont {Schwenk}},\ }\href
  {\doibase 10.1146/annurev-nucl-102313-025446} {\bibfield  {journal} {\bibinfo
   {journal} {Ann. Rev. Nucl. Part. Sci.}\ }\textbf {\bibinfo {volume} {65}},\
  \bibinfo {pages} {457} (\bibinfo {year} {2015})},\ \Eprint
  {http://arxiv.org/abs/1508.06893} {arXiv:1508.06893 [nucl-th]} \BibitemShut
  {NoStop}%
\bibitem [{\citenamefont {Tews}\ \emph {et~al.}(2018)\citenamefont {Tews},
  \citenamefont {Carlson}, \citenamefont {Gandolfi},\ and\ \citenamefont
  {Reddy}}]{Tews:2018kmu}%
  \BibitemOpen
  \bibfield  {author} {\bibinfo {author} {\bibfnamefont {I.}~\bibnamefont
  {Tews}}, \bibinfo {author} {\bibfnamefont {J.}~\bibnamefont {Carlson}},
  \bibinfo {author} {\bibfnamefont {S.}~\bibnamefont {Gandolfi}}, \ and\
  \bibinfo {author} {\bibfnamefont {S.}~\bibnamefont {Reddy}},\ }\href
  {\doibase 10.3847/1538-4357/aac267} {\bibfield  {journal} {\bibinfo
  {journal} {Astrophys. J.}\ }\textbf {\bibinfo {volume} {860}},\ \bibinfo
  {pages} {149} (\bibinfo {year} {2018})},\ \Eprint
  {http://arxiv.org/abs/1801.01923} {arXiv:1801.01923 [nucl-th]} \BibitemShut
  {NoStop}%
\bibitem [{\citenamefont {Hebeler}\ \emph {et~al.}(2010)\citenamefont
  {Hebeler}, \citenamefont {Lattimer}, \citenamefont {Pethick},\ and\
  \citenamefont {Schwenk}}]{Hebeler:2010jx}%
  \BibitemOpen
  \bibfield  {author} {\bibinfo {author} {\bibfnamefont {K.}~\bibnamefont
  {Hebeler}}, \bibinfo {author} {\bibfnamefont {J.~M.}\ \bibnamefont
  {Lattimer}}, \bibinfo {author} {\bibfnamefont {C.~J.}\ \bibnamefont
  {Pethick}}, \ and\ \bibinfo {author} {\bibfnamefont {A.}~\bibnamefont
  {Schwenk}},\ }\href {\doibase 10.1103/PhysRevLett.105.161102} {\bibfield
  {journal} {\bibinfo  {journal} {Phys. Rev. Lett.}\ }\textbf {\bibinfo
  {volume} {105}},\ \bibinfo {pages} {161102} (\bibinfo {year} {2010})},\
  \Eprint {http://arxiv.org/abs/1007.1746} {arXiv:1007.1746 [nucl-th]}
  \BibitemShut {NoStop}%
\bibitem [{\citenamefont {Chang}\ \emph {et~al.}(2017)\citenamefont {Chang},
  \citenamefont {Essig},\ and\ \citenamefont {McDermott}}]{Chang:2016ntp}%
  \BibitemOpen
  \bibfield  {author} {\bibinfo {author} {\bibfnamefont {J.~H.}\ \bibnamefont
  {Chang}}, \bibinfo {author} {\bibfnamefont {R.}~\bibnamefont {Essig}}, \ and\
  \bibinfo {author} {\bibfnamefont {S.~D.}\ \bibnamefont {McDermott}},\ }\href
  {\doibase 10.1007/JHEP01(2017)107} {\bibfield  {journal} {\bibinfo  {journal}
  {JHEP}\ }\textbf {\bibinfo {volume} {01}},\ \bibinfo {pages} {107} (\bibinfo
  {year} {2017})},\ \Eprint {http://arxiv.org/abs/1611.03864} {arXiv:1611.03864
  [hep-ph]} \BibitemShut {NoStop}%
\bibitem [{\citenamefont {Chang}\ \emph {et~al.}(2018)\citenamefont {Chang},
  \citenamefont {Essig},\ and\ \citenamefont {McDermott}}]{Chang:2018rso}%
  \BibitemOpen
  \bibfield  {author} {\bibinfo {author} {\bibfnamefont {J.~H.}\ \bibnamefont
  {Chang}}, \bibinfo {author} {\bibfnamefont {R.}~\bibnamefont {Essig}}, \ and\
  \bibinfo {author} {\bibfnamefont {S.~D.}\ \bibnamefont {McDermott}},\
  }\href@noop {} {\  (\bibinfo {year} {2018})},\ \Eprint
  {http://arxiv.org/abs/1803.00993} {arXiv:1803.00993 [hep-ph]} \BibitemShut
  {NoStop}%
\bibitem [{\citenamefont {Fornal}\ and\ \citenamefont
  {Grinstein}(2018)}]{Fornal:2018eol}%
  \BibitemOpen
  \bibfield  {author} {\bibinfo {author} {\bibfnamefont {B.}~\bibnamefont
  {Fornal}}\ and\ \bibinfo {author} {\bibfnamefont {B.}~\bibnamefont
  {Grinstein}},\ }\href {\doibase 10.1103/PhysRevLett.120.191801} {\bibfield
  {journal} {\bibinfo  {journal} {Phys. Rev. Lett.}\ }\textbf {\bibinfo
  {volume} {120}},\ \bibinfo {pages} {191801} (\bibinfo {year} {2018})},\
  \Eprint {http://arxiv.org/abs/1801.01124} {arXiv:1801.01124 [hep-ph]}
  \BibitemShut {NoStop}%
\bibitem [{\citenamefont {McKeen}\ \emph {et~al.}(2018)\citenamefont {McKeen},
  \citenamefont {Nelson}, \citenamefont {Reddy},\ and\ \citenamefont
  {Zhou}}]{McKeen:2018xwc}%
  \BibitemOpen
  \bibfield  {author} {\bibinfo {author} {\bibfnamefont {D.}~\bibnamefont
  {McKeen}}, \bibinfo {author} {\bibfnamefont {A.~E.}\ \bibnamefont {Nelson}},
  \bibinfo {author} {\bibfnamefont {S.}~\bibnamefont {Reddy}}, \ and\ \bibinfo
  {author} {\bibfnamefont {D.}~\bibnamefont {Zhou}},\ }\href {\doibase
  10.1103/PhysRevLett.121.061802} {\bibfield  {journal} {\bibinfo  {journal}
  {Phys. Rev. Lett.}\ }\textbf {\bibinfo {volume} {121}},\ \bibinfo {pages}
  {061802} (\bibinfo {year} {2018})},\ \Eprint
  {http://arxiv.org/abs/1802.08244} {arXiv:1802.08244 [hep-ph]} \BibitemShut
  {NoStop}%
\bibitem [{\citenamefont {Grinstein}\ \emph {et~al.}(2018)\citenamefont
  {Grinstein}, \citenamefont {Kouvaris},\ and\ \citenamefont
  {Nielsen}}]{Grinstein:2018ptl}%
  \BibitemOpen
  \bibfield  {author} {\bibinfo {author} {\bibfnamefont {B.}~\bibnamefont
  {Grinstein}}, \bibinfo {author} {\bibfnamefont {C.}~\bibnamefont {Kouvaris}},
  \ and\ \bibinfo {author} {\bibfnamefont {N.~G.}\ \bibnamefont {Nielsen}},\
  }\href@noop {} {\  (\bibinfo {year} {2018})},\ \Eprint
  {http://arxiv.org/abs/1811.06546} {arXiv:1811.06546 [hep-ph]} \BibitemShut
  {NoStop}%
\bibitem [{\citenamefont {Punturo}\ \emph {et~al.}(2010)\citenamefont {Punturo}
  \emph {et~al.}}]{Punturo:2010}%
  \BibitemOpen
  \bibfield  {author} {\bibinfo {author} {\bibfnamefont {M.}~\bibnamefont
  {Punturo}} \emph {et~al.},\ }\href
  {http://stacks.iop.org/0264-9381/27/i=19/a=194002} {\bibfield  {journal}
  {\bibinfo  {journal} {Classical and Quantum Gravity}\ }\textbf {\bibinfo
  {volume} {27}},\ \bibinfo {pages} {194002} (\bibinfo {year}
  {2010})}\BibitemShut {NoStop}%
\bibitem [{\citenamefont {Kouvaris}\ and\ \citenamefont
  {Tinyakov}(2010{\natexlab{b}})}]{Kouvaris:2010vv}%
  \BibitemOpen
  \bibfield  {author} {\bibinfo {author} {\bibfnamefont {C.}~\bibnamefont
  {Kouvaris}}\ and\ \bibinfo {author} {\bibfnamefont {P.}~\bibnamefont
  {Tinyakov}},\ }\href {\doibase 10.1103/PhysRevD.82.063531} {\bibfield
  {journal} {\bibinfo  {journal} {Phys. Rev.}\ }\textbf {\bibinfo {volume}
  {D82}},\ \bibinfo {pages} {063531} (\bibinfo {year} {2010}{\natexlab{b}})},\
  \Eprint {http://arxiv.org/abs/1004.0586} {arXiv:1004.0586 [astro-ph.GA]}
  \BibitemShut {NoStop}%
\bibitem [{\citenamefont {Kouvaris}\ and\ \citenamefont
  {Tinyakov}(2011)}]{Kouvaris:2010jy}%
  \BibitemOpen
  \bibfield  {author} {\bibinfo {author} {\bibfnamefont {C.}~\bibnamefont
  {Kouvaris}}\ and\ \bibinfo {author} {\bibfnamefont {P.}~\bibnamefont
  {Tinyakov}},\ }\href {\doibase 10.1103/PhysRevD.83.083512} {\bibfield
  {journal} {\bibinfo  {journal} {Phys. Rev.}\ }\textbf {\bibinfo {volume}
  {D83}},\ \bibinfo {pages} {083512} (\bibinfo {year} {2011})},\ \Eprint
  {http://arxiv.org/abs/1012.2039} {arXiv:1012.2039 [astro-ph.HE]} \BibitemShut
  {NoStop}%
\bibitem [{\citenamefont {Kouvaris}(2013)}]{Kouvaris:2013awa}%
  \BibitemOpen
  \bibfield  {author} {\bibinfo {author} {\bibfnamefont {C.}~\bibnamefont
  {Kouvaris}},\ }\href {\doibase 10.1155/2013/856196} {\bibfield  {journal}
  {\bibinfo  {journal} {Adv. High Energy Phys.}\ }\textbf {\bibinfo {volume}
  {2013}},\ \bibinfo {pages} {856196} (\bibinfo {year} {2013})},\ \Eprint
  {http://arxiv.org/abs/1308.3222} {arXiv:1308.3222 [astro-ph.HE]} \BibitemShut
  {NoStop}%
\bibitem [{\citenamefont {Cerme{\~n}o}\ \emph {et~al.}(2017)\citenamefont
  {Cerme{\~n}o}, \citenamefont {P{\'e}rez-Garc{\'\i}a},\ and\ \citenamefont
  {Silk}}]{Cermeno:2017xwb}%
  \BibitemOpen
  \bibfield  {author} {\bibinfo {author} {\bibfnamefont {M.}~\bibnamefont
  {Cerme{\~n}o}}, \bibinfo {author} {\bibfnamefont {M.~{\'A}.}\ \bibnamefont
  {P{\'e}rez-Garc{\'\i}a}}, \ and\ \bibinfo {author} {\bibfnamefont
  {J.}~\bibnamefont {Silk}},\ }\href {\doibase 10.1017/pasa.2017.38} {\bibfield
   {journal} {\bibinfo  {journal} {Publ. Astron. Soc. Austral.}\ }\textbf
  {\bibinfo {volume} {34}},\ \bibinfo {pages} {e043} (\bibinfo {year}
  {2017})},\ \Eprint {http://arxiv.org/abs/1710.06866} {arXiv:1710.06866
  [astro-ph.HE]} \BibitemShut {NoStop}%
\bibitem [{\citenamefont {Kopp}\ \emph {et~al.}(2018)\citenamefont {Kopp},
  \citenamefont {Laha}, \citenamefont {Opferkuch},\ and\ \citenamefont
  {Shepherd}}]{Kopp:2018jom}%
  \BibitemOpen
  \bibfield  {author} {\bibinfo {author} {\bibfnamefont {J.}~\bibnamefont
  {Kopp}}, \bibinfo {author} {\bibfnamefont {R.}~\bibnamefont {Laha}}, \bibinfo
  {author} {\bibfnamefont {T.}~\bibnamefont {Opferkuch}}, \ and\ \bibinfo
  {author} {\bibfnamefont {W.}~\bibnamefont {Shepherd}},\ }\href {\doibase
  10.1007/JHEP11(2018)096} {\bibfield  {journal} {\bibinfo  {journal} {JHEP}\
  }\textbf {\bibinfo {volume} {11}},\ \bibinfo {pages} {096} (\bibinfo {year}
  {2018})},\ \Eprint {http://arxiv.org/abs/1807.02527} {arXiv:1807.02527
  [hep-ph]} \BibitemShut {NoStop}%
\bibitem [{\citenamefont {Kaplan}\ \emph {et~al.}(2010)\citenamefont {Kaplan},
  \citenamefont {Krnjaic}, \citenamefont {Rehermann},\ and\ \citenamefont
  {Wells}}]{Kaplan:2009de}%
  \BibitemOpen
  \bibfield  {author} {\bibinfo {author} {\bibfnamefont {D.~E.}\ \bibnamefont
  {Kaplan}}, \bibinfo {author} {\bibfnamefont {G.~Z.}\ \bibnamefont {Krnjaic}},
  \bibinfo {author} {\bibfnamefont {K.~R.}\ \bibnamefont {Rehermann}}, \ and\
  \bibinfo {author} {\bibfnamefont {C.~M.}\ \bibnamefont {Wells}},\ }\href
  {\doibase 10.1088/1475-7516/2010/05/021} {\bibfield  {journal} {\bibinfo
  {journal} {JCAP}\ }\textbf {\bibinfo {volume} {1005}},\ \bibinfo {pages}
  {021} (\bibinfo {year} {2010})},\ \Eprint {http://arxiv.org/abs/0909.0753}
  {arXiv:0909.0753 [hep-ph]} \BibitemShut {NoStop}%
\bibitem [{\citenamefont {Tisserand}\ \emph {et~al.}(2007)\citenamefont
  {Tisserand} \emph {et~al.}}]{Tisserand:2006zx}%
  \BibitemOpen
  \bibfield  {author} {\bibinfo {author} {\bibfnamefont {P.}~\bibnamefont
  {Tisserand}} \emph {et~al.} (\bibinfo {collaboration} {EROS-2}),\ }\href
  {\doibase 10.1051/0004-6361:20066017} {\bibfield  {journal} {\bibinfo
  {journal} {Astron. Astrophys.}\ }\textbf {\bibinfo {volume} {469}},\ \bibinfo
  {pages} {387} (\bibinfo {year} {2007})},\ \Eprint
  {http://arxiv.org/abs/astro-ph/0607207} {arXiv:astro-ph/0607207 [astro-ph]}
  \BibitemShut {NoStop}%
\bibitem [{\citenamefont {Bowman}\ \emph {et~al.}(2018)\citenamefont {Bowman},
  \citenamefont {Rogers}, \citenamefont {Monsalve}, \citenamefont {Mozdzen},\
  and\ \citenamefont {Mahesh}}]{Bowman:2018fk}%
  \BibitemOpen
  \bibfield  {author} {\bibinfo {author} {\bibfnamefont {J.~D.}\ \bibnamefont
  {Bowman}}, \bibinfo {author} {\bibfnamefont {A.~E.~E.}\ \bibnamefont
  {Rogers}}, \bibinfo {author} {\bibfnamefont {R.~A.}\ \bibnamefont
  {Monsalve}}, \bibinfo {author} {\bibfnamefont {T.~J.}\ \bibnamefont
  {Mozdzen}}, \ and\ \bibinfo {author} {\bibfnamefont {N.}~\bibnamefont
  {Mahesh}},\ }\href {http://dx.doi.org/10.1038/nature25792} {\bibfield
  {journal} {\bibinfo  {journal} {Nature}\ }\textbf {\bibinfo {volume} {555}},\
  \bibinfo {pages} {67 EP } (\bibinfo {year} {2018})}\BibitemShut {NoStop}%
\bibitem [{\citenamefont {Barkana}(2018)}]{Barkana:2018uq}%
  \BibitemOpen
  \bibfield  {author} {\bibinfo {author} {\bibfnamefont {R.}~\bibnamefont
  {Barkana}},\ }\href {http://dx.doi.org/10.1038/nature25791} {\bibfield
  {journal} {\bibinfo  {journal} {Nature}\ }\textbf {\bibinfo {volume} {555}},\
  \bibinfo {pages} {71 EP } (\bibinfo {year} {2018})}\BibitemShut {NoStop}%
\bibitem [{\citenamefont {Randall}\ \emph {et~al.}(2008)\citenamefont
  {Randall}, \citenamefont {Markevitch}, \citenamefont {Clowe}, \citenamefont
  {Gonzalez},\ and\ \citenamefont {Bradac}}]{Randall:2007ph}%
  \BibitemOpen
  \bibfield  {author} {\bibinfo {author} {\bibfnamefont {S.~W.}\ \bibnamefont
  {Randall}}, \bibinfo {author} {\bibfnamefont {M.}~\bibnamefont {Markevitch}},
  \bibinfo {author} {\bibfnamefont {D.}~\bibnamefont {Clowe}}, \bibinfo
  {author} {\bibfnamefont {A.~H.}\ \bibnamefont {Gonzalez}}, \ and\ \bibinfo
  {author} {\bibfnamefont {M.}~\bibnamefont {Bradac}},\ }\href {\doibase
  10.1086/587859} {\bibfield  {journal} {\bibinfo  {journal} {Astrophys. J.}\
  }\textbf {\bibinfo {volume} {679}},\ \bibinfo {pages} {1173} (\bibinfo {year}
  {2008})},\ \Eprint {http://arxiv.org/abs/0704.0261} {arXiv:0704.0261
  [astro-ph]} \BibitemShut {NoStop}%
\end{thebibliography}
%

\end{document}